\documentclass[a4paper,12pt]{article}
\pdfoutput=1

\usepackage{jheppub}
\usepackage[utf8]{inputenc}
\usepackage{amsmath,amssymb,amsthm}
\usepackage{physics,mathtools}
\usepackage{tikz,float}
\usetikzlibrary{positioning}
\usepackage{bbm}
\usepackage[style=base]{subcaption} 
\usepackage{empheq}

\makeatletter
\def\@fpheader{\vspace{0.1mm}}
\makeatother
\newcommand{\bw}{\begin{widetext}}
\newcommand{\ew}{\end{widetext}}
\newcommand{\bea}{\begin{eqnarray}}
\newcommand{\eea}{\end{eqnarray}}
\newcommand{\be}{\begin{equation}}
\newcommand{\ee}{\end{equation}}
\newcommand{\bca}{\begin{cases}}
\newcommand{\eca}{\end{cases}}

\newcommand{\ben}{\begin{enumerate}}
\newcommand{\een}{\end{enumerate}}

\def\ie{\begin{equation}\begin{aligned}}
\def\fe{\end{aligned}\end{equation}}

\usetikzlibrary{patterns}
\makeatletter
\let\over\@@over
\makeatother

\title{
Thermal two-point functions in SYK and complex-time singularities
}

\author{Ilija Buri\'c${}^a$, Chi-Ming Chang${}^{b,c}$, Ivan Gusev${}^a$, Elizabeth Helfenberger${}^a$,\\ Andrei Parnachev${}^a$, Mukund Rangamani${}^d$}

\affiliation[a]{
School of Mathematics and Hamilton Mathematics Institute, \\
Trinity College, Dublin 2,
Ireland
}
\affiliation[b]{Yau Mathematical Sciences Center (YMSC), Tsinghua University, Beijing, China}

\affiliation[c]{Beijing Institute of Mathematical Sciences and Applications (BIMSA), Beijing, China}

\affiliation[d]{Center for Quantum Mathematics and Physics (QMAP)\\
Department of Physics \& Astronomy, University of California, Davis, CA 95616 USA}

\emailAdd{burici@tcd.ie,cmchang@tsinghua.edu.cn, gusevi@tcd.ie, helfenbe@tcd.ie, parnachev@maths.tcd.ie, mukund@physics.ucdavis.edu}

\abstract{
We analyze the finite-temperature two-point function of the large-$N$ SYK model at intermediate couplings away from the infrared fixed point. Specifically, we examine its analytic structure in the complex time plane, tracking the complex-time singularities over a range of temperatures. The location of the leading singularity lies on the imaginary axis. It controls the short-time dynamics of operator complexity, defining an `effective temperature' for the correlator. The next-to-leading singularity lies outside the thermal strip set by the above effective temperature. It has been argued that this could be interpreted in terms of bouncing null geodesics in the emergent black hole geometry. Both  these singularities persist all the way down to zero temperature. We discuss our observations and motivate the related emergent geometry  using a  kinematic space perspective.

}

\begin{document}

\maketitle

\section{Introduction}

The holographic gauge/gravity correspondence has been instrumental at eliciting features of strongly coupled quantum field theories. Computation of observables is facilitated by the fact that the strong coupling large $N$ dynamics of such field theories is geometrized by a classical gravitational dynamics. For example, correlation functions of gauge invariant operators can be boiled down to solving classical differential equations. Consider for instance, the retarded Green's function at finite temperature: the holographic computation makes clear the analytic structure. The frequency space correlator is meromorphic, indicating late time relaxation to thermal equilibrium, and can be understood to be a direct consequence of quasinormal modes of the dual black hole geometry~\cite{Horowitz:1999jd}.\footnote{In contrast, a weakly coupled thermal field theory exhibits branch cuts in frequency space~\cite{Hartnoll:2005ju}.}. We will be interested in this observable (and its cousins) in simple quantum mechanical models. 

The analytic structure in frequency domain has been well understood thanks to the fact that wave equations in stationary black hole backgrounds are best solved by Fourier mode decomposition. On the other hand, the picture for the time-domain correlator has only become clear relatively recently. The Euclidean two-point function is required to be analytic in the fundamental strip $\tau \in (0, \beta)$, where $\beta$ is the inverse temperature and $\tau$ is Euclidean time. 
Together with the Kubo-Martin-Schwinger (KMS) condition, analyticity in this strip is the Euclidean counterpart of the fluctuation-dissipation relation for the real-time correlator.
What happens outside the strip depends on the observable --- in quantum field theories the correlator $\expval{\mathcal{O}(t,\vb{x}) \, \mathcal{O}(0,0)}_\beta$ with fixed spatial separation possesses light-cone cuts originating at $\operatorname{Re}(t) = \pm|{\vb{x}}|$ on the lower edge of the strip, $\operatorname{Im}(t) = 0$, together with their KMS images at $\operatorname{Im}(t) = \beta$. 
If we consider spatially smeared correlators, say by Fourier transforming to momentum space, then at fixed spatial momentum the edge of the thermal strip becomes a true boundary of the domain of analyticity. Such smeared correlators might then have additional singularities outside the thermal strip. In particular, in strongly coupled quantum mechanical models (with a suitable large $N$ limit), absent the ability to spatially separate operators, one should anticipate their presence.  

A further motivation for understanding the time-domain singularities of thermal observables comes from the geometry. For operators of high dimension, one may motivate an eikonal approximation for the wave equation, leading to the study of null geodesics. Doing so, one can argue for time-domain singularities from the classical black hole geometry arising directly from the curvature singularity of the black hole~\cite{Fidkowski:2003nf} (see also~\cite{Louko:2000tp,Kraus:2002iv} for earlier attempts using geodesic probes to study black hole interiors). In the geodesic approximation, this time-domain singularity originates from a pair of distinguished null geodesics, dubbed the `bouncing null geodesic' in the literature, that picks out a characteristic timescale $t_c$ outside the thermal strip. The origin of the time-domain singularity may equivalently be traced directly to the asymptotic behavior of highly damped quasinormal modes~\cite{Festuccia:2005pi}. This picture has become clear from  the position space
\cite{Ceplak:2024bja,Ceplak:2025dds,Araya:2026shz}
and momentum space analysis of thermal correlators \cite{Afkhami-Jeddi:2025wra,Jia:2025jbi,AliAhmad:2026wem,Giombi:2026kdz,Jia:2026ryl}
(see also \cite{Parisini:2023nbd,Jia:2026pmv,Arnaudo:2026der,Grozdanov:2026cut,Grozdanov:2026ktq} for related work).

Most of the aforementioned works infer the properties of thermal correlators and the imprint of the black hole singularity thereupon using the gravitational description. A complementary and useful exercise is to explicitly work out the same starting from the field theory itself. While strongly coupled planar gauge theories are presently out of reach, large $N$ quantum mechanical models provide a viable avenue for this analysis. The paradigmatic model that retains tractability while capturing interesting physical insights is the Sachdev–Ye–Kitaev (SYK) model~\cite{Sachdev:1992fk,KitaevTalks,Maldacena:2016hyu,Kitaev:2017awl}. Indeed, it was shown that the infinite-temperature SYK model at weak coupling exhibits discrete quasinormal modes in the thermal spectrum~\cite{Dodelson:2024atp}. The asymptotic behavior of these modes at high damping lead to time-domain singularities of the form seen in the black hole examples. Furthermore, it was argued that one  can mimic finite string length corrections to gravitational dynamics by using the degree of fermion coupling in the Hamiltonian as a dial~\cite{Dodelson:2025jff} (see below). 

In this paper, we will go beyond these analyses, and extend~\cite{Dodelson:2025jff} to study singularities of the large-$N$ SYK two-point function at finite (rather than infinite) temperature. We recall that the SYK model comprises $N$ Majorana fermions with a suitable all-to-all interaction Hamiltonian built out of $q$-body terms. The couplings are drawn from a random Gaussian ensemble, and it is this disorder averaging that makes the model tractable. Despite its simplicity, the model exhibits a remarkable combination of solvability and rich dynamical behavior. In the infrared, its large-$N$ dynamics are governed by an emergent approximate conformal symmetry and are holographically described by nearly-AdS$_2$ Jackiw–Teitelboim (JT) gravity~\cite{Jensen:2016pah,Maldacena:2016upp}. In this regime, it also saturates the Maldacena–Shenker–Stanford bound on quantum chaos \cite{Maldacena:2016hyu,Maldacena:2015waa}, a necessary condition for a theory with a gravity dual. This puts the holographic description of SYK in the infrared (IR) on firm footing. However, much less is known about the holographic interpretation of its mid-energy range and ultraviolet (UV) dynamics (see \cite{Zhang:2020jhn} for a recent discussion). It is therefore of great interest to explore SYK observables away from the IR limit, which will be our focus. Note that the  $q\to \infty$ limit has a simple description~\cite{Cotler:2016fpe} even away from the IR critical point. This was exploited in~\cite{Dodelson:2025jff} to argue that $\frac{1}{q}$ could be viewed as a proxy for the string tension, with $q\to \infty$ akin to the tensionless limit. 
\smallskip

In any regime, SYK admits a controlled large-$N$ expansion in which physical observables can be computed systematically, thanks to the truncation of the Schwinger-Dyson (SD) equations (owing to disorder averaging). For the thermal two-point function, this implies that one can iteratively solve for the Green's function and the self-energy. We shall use two complementary methods to solve the finite temperature SD equations for the thermal two-point function. The first is a direct numerical solution of the system, while the second is a perturbative \emph{double-expansion} scheme. The latter exploits the fact that at finite temperature and coupling $J$, the correlator is a function of two dimensionless combinations 
$\tau/\beta$ and $\beta J$. The double-expansion is performed around the free UV point and complements the conformal limit that emerges in the IR. Large order behavior of coefficients, inferred from explicit computations, suggests that the radius of convergence in both variables is finite. In particular, for $\beta=0$, the radius of convergence of the $\tau J$ expansion is related to the universal operator growth hypothesis~\cite{Parker:2018yvk}. For low temperatures and/or long times, the numerical approach is of course more reliable. In the regime where both numerical and perturbative approaches apply, we will demonstrate that they agree very well with one another.
\smallskip  

Based on these two approaches, we show that at any temperature, the two-point function has an infinite number of singularities in the complex $\tau$-plane. We will track the positions of the leading few singularities as the temperature is varied from infinity towards zero. Curiously, they exhibit a relatively mild temperature dependence and in fact appear to persist all the way down to zero temperature! The leading singularity is interpreted as an `effective temperature' in the model, and in particular, guarantees that the correlator is analytic in a wider region than naively required by thermal KMS invariance. The next-to-leading singularity controls the high frequency behavior of the spectral density, and is reminiscent of the time domain singularity arising from bouncing null geodesics in holographic black hole backgrounds.
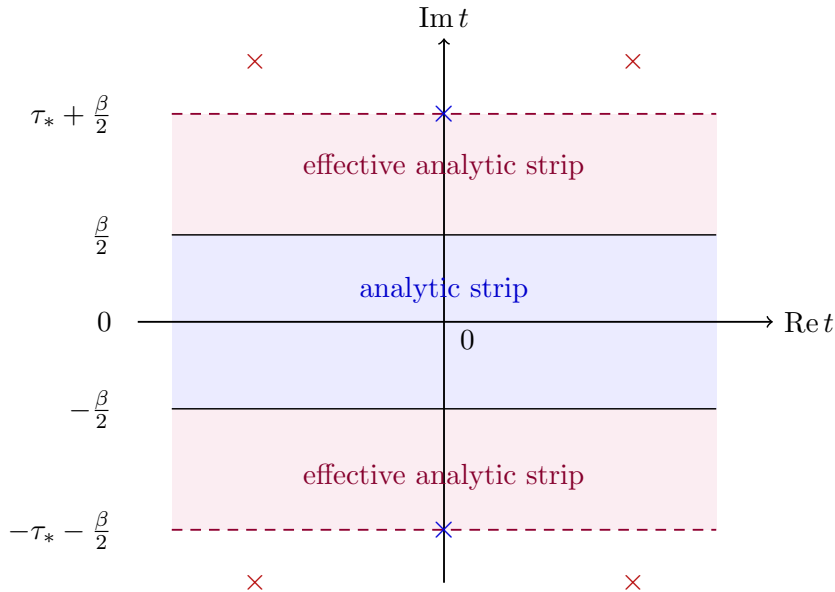
\begin{figure}[t]
\centering
\begin{tikzpicture}[
    x=1cm,y=1cm,
    every node/.style={font=\small},
    axis/.style={->, line width=0.7pt},
    boundary/.style={line width=0.55pt},
    effboundary/.style={dash pattern=on 4pt off 3pt, line width=0.7pt, purple!75!black},
    lead/.style={blue!85!black, line width=1.1pt},
    bounce/.style={red!70!black, line width=0.9pt},
]

\def\xL{-3.6}
\def\xR{3.6}
\def\xAxisL{-4.05}
\def\xAxisR{4.35}

\def\yTop{2.75}
\def\yHalf{1.15}
\def\yBot{-2.75}
\def\ymHalf{-1.15}

\def\xb{2.5}
\def\yb{3.45}

\fill[purple!7] (\xL,\yHalf) rectangle (\xR,\yTop);
\fill[blue!8]   (\xL,\ymHalf) rectangle (\xR,\yHalf);
\fill[purple!7] (\xL,\yBot) rectangle (\xR,\ymHalf);

\draw[effboundary] (\xL,\yTop) -- (\xR,\yTop);
\draw[effboundary] (\xL,\yBot) -- (\xR,\yBot);

\draw[boundary] (\xL,\yHalf) -- (\xR,\yHalf);
\draw[boundary] (\xL,\ymHalf) -- (\xR,\ymHalf);

\draw[axis] (0,-3.45) -- (0,3.75) node[above] {$\operatorname{Im} t$};
\draw[axis] (\xAxisL,0) -- (\xAxisR,0) node[right] {$\operatorname{Re} t$};

\node[below right=-1pt and 2pt] at (0,0) {$0$};

\node[lead, scale=1.15] at (0,\yTop) {$\times$};
\node[lead, scale=1.15] at (0,\yBot) {$\times$};

\node[bounce, scale=1.05] at (-\xb,\yb) {$\times$};
\node[bounce, scale=1.05] at ( \xb,\yb) {$\times$};
\node[bounce, scale=1.05] at (-\xb,-\yb) {$\times$};
\node[bounce, scale=1.05] at ( \xb,-\yb) {$\times$};

\node[left] at (-4.25,\yTop) {$\tau_*+\frac{\beta}{2}$};
\node[left] at (-4.25,\yHalf) {$\frac{\beta}{2}$};
\node[left] at (-4.25,0) {$0$};
\node[left] at (-4.25,\ymHalf) {$-\frac{\beta}{2}$};
\node[left] at (-4.25,\yBot) {$-\tau_*-\frac{\beta}{2}$};

\node[purple!70!black] at (0,2.05) {effective analytic strip};
\node[blue!80!black]   at (0,0.42) {analytic strip};
\node[purple!70!black] at (0,-2.05) {effective analytic strip};

\end{tikzpicture}

\caption{
Schematic analytic structure of the two-sided thermal correlator in the complex time plane.
Blue crosses denote the leading imaginary-time singularities, while red crosses denote the next-to-leading singularities.
}
\label{fig:two-sided-analytic-strip-bouncing}
\end{figure}
The analytic structure of the two-sided thermal correlator is shown in Figure \ref{fig:two-sided-analytic-strip-bouncing}.
In terms of implementation, we will always work at fixed finite $q$, with numerical data presented using $q=4$ for specificity.

The paper is organized as follows. In Section \ref{S:Finite temperature two-point function in SYK}, we review some known facts about the SYK model and present the double-expansion and numerical algorithms for computation of the large-$N$ two-point function. Section \ref{S:Singularities of the two-point function} presents the results of these two methods, and, in particular, the singularities in the complex time domain.
We discuss our results in Section \ref{S:Discussion}.

\noindent
\emph{Note added:} As this work was in progress we received \cite{Dodelson:2026gak} which has a high degree of overlap with our analysis. We focus on the analytic structure of the correlators in the complex time domain, while the aforementioned paper examines the  related quasinormal spectrum in frequency space.

\section{Finite temperature two-point function in SYK}
\label{S:Finite temperature two-point function in SYK}

We begin with a short review of the salient features of the SYK model in the first subsection, before describing two methods for solving the Schwinger-Dyson equations to obtain the thermal two-point function in the second and third subsection.

\subsection{The model}
 
The SYK model is a (0+1)-$d$ quantum-mechanical system consisting of $N$ Majorana fermions with all-to-all $q$-particle random interactions, where $q$ is an even positive integer. Its Hamiltonian reads
\begin{equation}\label{SYK-Hamiltonian}
    H = i^{\frac{q}{2}} \sum_{1\leq i_1\leq i_2\leq\ldots\leq i_q\leq N} J_{i_1 i_2\ldots i_q}\psi_{i_1}\psi_{i_2}\ldots\psi_{i_q}\ .
\end{equation}
Here, $\psi_i$ are Majorana fermions satisfying canonical anti-commutation relations
\begin{equation}\label{canonical-anticommutators}
    \{\psi_i,\psi_j\} = \delta_{ij}\,,
\end{equation}
and the $N \choose q$ couplings $J_{i_1 i_2\ldots i_q}$ are independent identically distributed random variables drawn from a Gaussian distribution with zero mean and variance given by
\begin{equation}\label{J}
\langle J_{i_1 i_2\ldots i_q}^2\rangle = \frac{J^2(q-1)!}{N^{q-1}} = \frac{2^{q-1}}{q} \frac{\mathcal{J}^2}{N^{q-1}}\ .
\end{equation}
Interactions in the model are characterized by the dimension one parameter $J$ (equivalently, $\mathcal{J}$, which is defined to allow uniform behavior as a function $q$). The Hamiltonian is Hermitian for even $q$. Throughout the paper, we work in the strict $N\to\infty$ limit.
\smallskip

We are interested in the two-point function of the fundamental fermion at finite temperature $T = \beta^{-1}$,
\begin{equation}\label{Euclidean-time-ordered-main}
    G(\tau) = \langle\psi(\tau)\psi(0)\rangle_\beta\, \theta(\tau) - \langle\psi(0)\psi(\tau)\rangle_\beta\, \theta(-\tau)\ .
\end{equation}
Here $\tau$ is the Euclidean time, $G(\tau)$ the time-ordered Euclidean propagator, and the thermal expectation value is defined as
\begin{equation}
   \langle \psi(\tau)\psi(0) \rangle_\beta \equiv \frac{\text{tr} \left(\psi(\tau) \psi(0) e^{-\beta H}\right)}{\text{tr}\left(e^{-\beta H}\right)}\ .
\end{equation}
Its relation to Wightman and retarded two-point functions is reviewed in Appendix \ref{A:Two-point functions at finite temperature}. At large $N$, the two-point function \eqref{Euclidean-time-ordered-main} satisfies the Schwinger-Dyson (SD) equations~\cite{Maldacena:2016hyu},
\begin{equation}\label{SYK SDE Euclidian}
    G^{-1}(\omega_n) = -i\omega_n - \Sigma(\omega_n)\,, \qquad 
    \Sigma(\tau) = J^2 \,G^{q-1}(\tau)\,,     
\end{equation}
where $\omega_n$ are fermionic Matsubara frequencies
\begin{equation}\label{fermion-frequencies-main}
    \omega_n = \frac{2\pi}{\beta}\left(n+\frac12\right)\,, \qquad n\in\mathbb{Z}\ .
\end{equation}
The equations involve both frequency and time domain functions --- the Green's function equation is written in frequency space while the self-energy is determined in the time domain. These equations also hold at zero temperature; one simply replaces  $\omega_n$ by a real variable $\omega$ and extends $\tau$ from $[0,\beta)$ to the whole real line. At zero coupling, $J=0$, the two-point function reduces to
\begin{equation}\label{time-Euclidean-free}
    G^{(0)}(\tau) = \frac12 \text{sgn}(\tau)\,, \qquad G^{(0)}(\omega_n) = \frac{i}{\omega_n}\ .
\end{equation}
This holds in the fundamental domain $\tau\in[0,\beta)$. Using the thermal KMS symmetry, one may  periodically extend the result to the real $\tau$-line using $G^{(0)}(\tau+\beta) = - G^{(0)}(\tau)$. At finite coupling $J$, the propagator is a non-trivial function of two dimensionless parameters, $\tau/\beta$ and $\beta J$. It also depends on the parameter $q$. For most of our discussion, we will keep $q=4$ and determine the propagator as a function of $(\tau/\beta,\beta J)$ in various regimes of the two arguments. We begin by reviewing the two known limits: the conformal limit and the infinite temperature limit.

\subsection{Conformal and infinite temperature limits}

Since the two-point function \eqref{Euclidean-time-ordered-main} depends on the dimensionless quantities $(\tau/\beta,\beta J)$, we shall, without loss of generality, set $J=1$ for the remainder of this section and write $G = G_\beta(\tau)$.
\smallskip

The conformal limit is given by $\beta,\tau\to\infty$ with $\tau/\beta$ finite. In this regime, the two-point function is given by (see e.g. \cite{Maldacena:2016hyu})
\begin{equation}\label{conformal-answer-finite-T}
    G_c(\tau) = b \left( \frac{\pi}{\beta \sin\frac{\pi\tau}{\beta}}\right)^{2/q} \text{sgn}(\tau)\,,
\end{equation}
where the normalization constant $b$ is given as the solution to the equation
\begin{equation}
    b^q = \frac{q-2}{2\pi\,q} \,\tan\frac{\pi}{q}\ .
\end{equation}
If we further assume $\tau\ll\beta$, the two-point function \eqref{conformal-answer-finite-T} reduces to the flat space conformal correlator
\begin{equation}\label{conformal-answer-zero-T}
    G_c^{\tau\ll\beta}(\tau) = \frac{b}{|\tau|^{2/q}}\, \text{sgn}(\tau)\ .
\end{equation}
Another well understood regime is the short-time limit at infinite temperature. Here we must work in Lorentzian time $t=-i\tau$, as the thermal circle collapses to zero. In this regime, it is possible to solve the SD equations order by order in the small $t$ expansion \cite{Parker:2018yvk,Dodelson:2024atp}. This expansion has a finite radius of convergence. To go beyond this radius, one may truncate the series and use a Pad\'e approximation. This method was used in \cite{Dodelson:2025jff} to predict the positions of correlator singularities in the complex $t$-plane. Since we will use a similar approach for the finite temperature case\footnote{The finite temperature dynamics of operator growth in SYK model was investigated in \cite{Qi:2018bje}.} below, we present an overview of the general scheme.

For real-time dynamics, one typically works with the retarded correlation function $G_R (t)$ rather than the time-ordered one, since it captures the causal response of the system to a perturbation. The fermionic retarded Green's function is defined as 
\begin{equation}
    G_R (t) \equiv \theta(t) \langle\{\psi(t), \psi(0) \} \rangle_\beta\ .
\end{equation}
It will be convenient to write $G_R(t)$ in terms of the following greater/lesser Wightman functions\footnote{This can be obtained by analytic continuation of the time-ordered Euclidean correlator as described in Appendix \ref{A:Two-point functions at finite temperature}.} 
\begin{equation}
        G^> (t) \equiv \langle \psi(t) \psi(0) \rangle_\beta\,, \qquad 
        G^< (t) \equiv \langle \psi(0) \psi(t) \rangle_\beta\ . 
\end{equation}
Then $G_R(t)$ is given by
\begin{equation}
    G_R(t) = \theta(t) (G^> (t) + G^<(t))\ . 
\end{equation}

\paragraph{Infinite temperature expansion algorithm:}

The main idea of the method is to expand the two-point function in a power series in $t$, compute the corresponding expansion of the self-energy and then impose the Schwinger-Dyson equation order by order at large frequency.

The real-time Schwinger-Dyson equation at infinite temperature can be written as (see Appendix \ref{A: Analytic continuation}, equation \eqref{real-time SDE})
\begin{align}\label{real-time-SDE-inf-temp}
    &G_R^{-1}(\omega) = -i\omega + \varepsilon -i\Sigma(-i\omega + \varepsilon)\,, \nonumber \\ 
    &\Sigma(-i\omega+\varepsilon)\Big|_{\beta=0} = 2i\, \int_0^\infty dt\; e^{i\omega t-\varepsilon t}\,\big(G^>(t)\big)^{q-1}\,, \\ \nonumber
&G^>(\omega) = \text{Re}\ G_R(\omega)\,, \qquad \varepsilon\to0\ .
\end{align}
The expansion coefficients in the $t$-series, i.e., the moments, will be denoted by $\mu_n$. More precisely, we write the expansion of the two-point function as
\begin{equation}\label{Gt expansion}
G^>(t) = \sum_{n=0}^\infty \mu_n  \frac{(it)^{2n}}{(2n)!}\,, \qquad \mu_0=\frac{1}{2}\ .
\end{equation}
We have used the fact that, at infinite temperature, only even powers of $t$ appear. Our aim is to solve for the coefficients $\mu_n$. Since the self-energy involves the power $G^>(t)^{q-1}$, we also need to define expansion coefficients $Q_n$ as
\begin{equation}\label{Gtp expansion}
G^>(t)^{q-1} = \sum_{n=0}^\infty Q_n\frac{(it)^{2n}}{(2n)!}\ .
\end{equation}
Explicitly, using binomial expansion for the left hand side 
\begin{equation}
    Q_n = (2n)!\sum_{\substack{k_1+\cdots+k_{q-1}=n \\ k_i\ge 0}}\ 
\prod_{i=1}^{q-1}
\frac{\mu_{k_i}}{(2k_i)!}\,,
\end{equation}
with $Q_0 = \frac{1}{2^{q-1}}$. To construct the algorithm, we need to determine the moment $\mu_n$ from the data $\mu_0\,,\ldots\,, \mu_{n-1}$. To this end, we expand the two-point function in frequency space
\begin{equation}
      G_R(\omega) =  2i\sum_{n=0}^\infty \frac{\mu_n}{\omega^{2n+1}}\ .
\end{equation}
The corresponding expansion of the self-energy reads
\begin{equation}
\Sigma(-i\omega+\varepsilon) = -2\sum_{n=0}^\infty \frac{Q_n}{\omega^{2n+1}}\ .
\end{equation}
Finally, we substitute the last two expansions in the Schwinger-Dyson equation
\begin{equation}\label{rec Dyson}
(-i\omega)G_R(\omega) = 1+ i\Sigma(-i\omega+\varepsilon)G_R(\omega)\ .
\end{equation}
Equating the coefficients with different powers of $\omega$ gives
\begin{equation}\label{recursion-final-zero-temp}
   \mu_n = 2\sum_{k=0}^{n-1} Q_k\, \mu_{n-1-k}\,, \qquad n\geq 1\,,
\end{equation}
where, as expected, the right hand side depends only on lower moments $\mu_0\,,\ldots\,, \mu_{n-1}$. This completes the specification of the iterative solution algorithm. 

\subsection{Double-expansion perturbative solution to SD equations}

We now turn to the perturbative solution of the Schwinger-Dyson equations at finite temperature. Here, it will be more convenient to work in Euclidean time $\tau$. We introduce a dimensionless time coordinate 
\begin{equation}
    u=\frac{\tau}{\beta}\,, \qquad0 < u < 1\,, 
\end{equation}
and write the finite-temperature expansion in the form
\begin{equation}\label{finite-beta-G-expansion}
G(\tau) = \sum_{n=0}^\infty \beta^{2n} m_n(u)\,, \qquad m_0(u)=\frac12\ .
\end{equation}
For the self-energy, we define the coefficients $\sigma_n(u)$ by
\begin{equation}\label{sigma-expansion}
\beta^2\Sigma(\tau) = \sum_{n=1}^\infty \beta^{2n}\sigma_n(u)\ .
\end{equation}
Explicitly, using the second SD equation in \eqref{SYK SDE Euclidian} and the binomial expansion for the left hand side, we have
\begin{equation}
    \sigma_n(u) = \sum_{\substack{k_1+\cdots+k_{q-1}=n-1\\ k_i\ge 0}}\ \prod_{j=1}^{q-1}m_{k_j}(u)\,, \qquad n\ge 1\ .
\end{equation}
Therefore, the coefficients $\sigma_n(u)$ are completely determined once $m_0(u),\ldots,m_{n-1}(u)$ are known. 

To solve the SD equations we pass over to frequency space. Let the  dimensionless Matsubara frequency be denoted by $\Omega=\beta\omega_m$. The Green's function is expanded as
\begin{equation}\label{finite-beta-G-frequency-expansion}
\frac{1}{\beta}\, G(\omega) = \sum_{n=0}^\infty \beta^{2n}g_n(\Omega)\,, \qquad g_0(\Omega)=\frac{i}{\Omega}\ .
\end{equation}
Similarly, we write
\begin{equation}\label{finite-beta-Sigma-frequency-expansion}
\beta\Sigma(\omega) = \sum_{n=1}^{\infty} \beta^{2n} s_n(\Omega)\ .
\end{equation}
The coefficients $g_n(\Omega)$ and $s_n(\Omega)$ are the finite-temperature Fourier transforms of $m_n(u)$ and $\sigma_n(u)$,
\begin{equation}\label{sn-Fourier}
g_n(\Omega) = \int_0^1 du\,
e^{i\Omega u}m_n(u)\,, \qquad s_n(\Omega) = \int_0^1 du\,
e^{i\Omega u}\sigma_n(u)\ .
\end{equation}
Since $\sigma_n(u)$ is a polynomial in $u$, the Fourier transform gives a finite large-frequency expansion in powers of $1/\Omega$; see Appendix~\ref{useful formulas} for details. Finally, we impose the Schwinger-Dyson equation
\begin{equation}\label{finite-beta-Dyson}
(-i\omega)G(\omega) = 1+\Sigma(\omega)G(\omega)\,,
\end{equation}
which, upon using~\eqref{finite-beta-G-frequency-expansion} and~\eqref{finite-beta-Sigma-frequency-expansion}, gives
\begin{equation}\label{finite-beta-g-recursion}
g_n(\Omega) = \frac{i}{\Omega} \sum_{r=1}^n s_r(\Omega)g_{n-r}(\Omega)\,, \qquad n\geq 1\ .
\end{equation}
This determines $g_n$ from the previously computed data. The final step is to transform $g_n(\Omega)$ back to Euclidean time and rewrite the result as a polynomial in $u$, obtaining thus the next coefficient $m_n(u)$. 

Carrying out the above procedure to a few leading orders, we find 
\begin{equation}
\begin{aligned}
G(u)& = \ \frac12+\frac{\beta^2}{16}\,u(u-1) +\frac{\beta^4}{192}\,u\,(u-1)\,(u^2-u-1)\\
&+\frac{\beta^6}{92160}\,u\,(u-1) \left(37\,u^4-74\,u^3-29\,u^2+66\,u+66\right) +\mathcal O(\beta^8)\ .
\end{aligned}
\end{equation}
We can reorganize the expansion by collecting powers of $\tau$, viz., 
\begin{equation}\label{G tau beta}
\begin{aligned}
G(\tau)
=&\ \frac12 +\left(-\frac{\beta}{16}+\frac{\beta^3}{192}-\frac{66\,\beta^5}{92160} + \mathcal{O}(\beta^7)\right)\tau
 + \left(\frac{1}{16}+\mathcal{O}(\beta^2)\right)\tau^2\\
 +& \left(\frac{95\,\beta^3}{92160}+\mathcal{O}(\beta^5)\right) \tau^3 + \left(\frac{1}{192}+\frac{45\,\beta^2}{92160} + \mathcal{O}(\beta^4)\right)\tau^4
+\mathcal O(\tau^5)\ .
\end{aligned}
\end{equation}
This representation allows us to take the limit $\beta\to0$, $\tau=it$
\begin{equation}
    \left.G(\tau)\right|_{\beta=0} = \frac12-\frac{t^2}{16}+\frac{t^4}{192}-\frac{37\,t^6}{92160}+\mathcal O(t^8)\,,
\end{equation}
which agrees with the expansion of the two-point function found in \cite{Dodelson:2024atp}.

Note that each iteration of the algorithm preserves KMS invariance and Hermiticity of the solution. The KMS condition acts as a global constraint, much like in recent thermal bootstrap frameworks \cite{Buric:2025anb,Buric:2025fye,Niarchos:2025cdg,Barrat:2025twb}, and uniquely fixes the coefficients in the double expansion. Each time moment, as a series in $\beta$, has a finite radius of convergence in the complex $\beta$-plane at fixed $\tau$. Numerically, we find that the large-order behavior of the coefficients is controlled by singularities located at
\begin{equation}
    \beta=\pm 2i\,,
\end{equation}
so that the domain of convergence is $|\beta|<2$. The finite radius of convergence of this expansion in fact has a simple origin. It follows from the fermionic thermal factor. After analytic continuation to complex $\beta$, the Fermi-Dirac distribution becomes singular when
\begin{equation}
1+e^{-\beta\omega_n}=0
\qquad \Longleftrightarrow \qquad
\beta\omega_n=i\pi(2m+1)\,, \qquad m\in\mathbb Z\ .
\end{equation}
The closest such singularities are originated from $n=m=0$ and located at $\beta=\pm 2i$. 

Our ultimate goal is to study the correlator for temperatures extending well beyond this domain. For this purpose, the power series by itself is insufficient. To analytically continue beyond the convergence disk, we resum the $\beta$-series using Pad\'e approximants. More precisely, for every fixed value of $u$, we replace the truncated series by a diagonal Pad\'e approximant in the variable $\beta^2$. 

At this point, it is worth differentiating two different Pad\'e approximations which enter our analysis. The first is the Pad\'e resummation in $\beta^2$ described above, which analytically continues the high-temperature expansion to finite and low temperatures. Having done this, we perform an additional Pad\'e approximation in the Euclidean time variable $\tau$. This is used to probe the correlator's intermediate-time analytic structure, including the poles in the complex $t$-plane, as discussed in the next section.

\subsection{Numerical solution to SD equations}

Finally, let us describe the numerical algorithm for solving the SD equations. We use a damped fixed-point iteration procedure, following the strategy described in~\cite[ Appendix~G]{Maldacena:2016hyu}. In this method, one iterates the two-point function and uses a weighted update to improve convergence. We will write the algorithm for computing the Wightman correlator at real time. A similar procedure applies to the Euclidean correlator.
\smallskip

We truncate and discretize both the time and the frequency axes, so that all the functions, either in time or frequency domains are replaced by arrays of $\Lambda^2$ elements. Here, $\Lambda$ is a large number fixed at the outset, to be thought of as the ‘UV cutoff'. To avoid confusion, we denote the discrete two-point functions by $G$ in the time domain and by $\widetilde{G}$ in the frequency domain. The algorithm also passes through the spectral density $\rho$ and the self-energy $\Sigma$ (these will always appear only in the frequency domain in intermediate steps). Finally, let us agree to write $\mathcal{W}$ for the array
\begin{equation}
    \mathcal{W} \equiv  \left(2\pi \left( i +\frac12 \right)\,, \quad i = -\frac{\Lambda^2}{2},-\frac{\Lambda^2}{2}+1,\dots,\frac{\Lambda^2}{2}-1 \right)\ .
\end{equation}
With this notation in place, we can spell out the algorithm. Starting with the initial condition
\begin{equation}
    \Sigma^{(0)} = i\,,
\end{equation}
we iteratively compute $\Sigma^{(n)}$, $\widetilde{G}_R^{(n)}$, $\rho^{(n)}$ and $G^{\lessgtr,(n)}$ in the following order  
\begin{equation}
    \Sigma^{(n)} \longrightarrow \widetilde{G}_R^{(n+1)} \longrightarrow  \rho^{(n+1)} \longrightarrow G^{\lessgtr,(n+1)} \longrightarrow \Sigma^{(n+1)}\ .
\end{equation}
The spectral density and the Wightman functions $G^{\lessgtr}$ are useful intermediaries for organizing the computation.  
The individual steps, dictated by SD equations, read\footnote{These are just the discrete versions of the SD equations \eqref{real-time SDE} and the familiar relations
\begin{equation}
    \rho(\omega) = 2 \,\text{Re} G_R(\omega) = \left(1+e^{-\beta\omega}\right) G^>(\omega) = \left(1+e^{\beta\omega}\right) G^<(\omega)\,,
\end{equation}
between the spectral density and Wightman and retarded Green's function. See Appendices \ref{A:Two-point functions at finite temperature} and \ref{A: Analytic continuation} for more details.}
\begin{align}
    & \widetilde{G}_R^{(n+1)} = \frac{1}{-i\mathcal{W}-i\Sigma^{(n)}}, \hskip1.9cm  \rho^{(n+1)} = 2\,\mathrm{Re}\, \widetilde{G}_R^{(n+1)}\,,\\
    & G^{>,(n+1)} = \mathcal{F}^{-1}\!\left[\frac{\rho^{(n+1)}}{1+e^{-\beta\mathcal{W}}}\right]\,, \qquad
 G^{<,(n+1)} = \mathcal{F}^{-1}\!\left[\frac{\rho^{(n+1)}}{1+e^{\beta\mathcal{W}}}\right]\,,
\end{align}
where $\mathcal{F}^{-1}$ denotes the discrete inverse Fourier transform on the above chosen grid. Finally, the self-energy is updated through
\begin{equation}
\Sigma_{\rm raw}^{(n+1)} =
iJ^2\,\mathcal{F}\!\left[
\Theta(t)\Big((G^{>,(n+1)})^{q-1}+(G^{<,(n+1)})^{q-1}\Big)\right],
\end{equation}
followed by a damped relaxation step
\begin{equation}
\Sigma^{(n+1)} = (1-\lambda_n)\Sigma^{(n)} + \lambda_n\,\Sigma_{\rm raw}^{(n+1)}\ .
\end{equation}
The last step requires some additional explanation. The algorithm contains one more parameter $\lambda_n$, which gets updated in each iteration cycle. We start off with $\lambda_0=1/2$. In each cycle, we compute the quantity
\begin{equation}
    \text{increment}^{(n)} = \left|\widetilde{G}^{(n)}_R - \widetilde{G}^{(n-1)}_R\right|_1 + \frac{1}{\lambda_{n-1}} \left|\Sigma^{(n)} - \Sigma^{(n-1)}\right|_1\,,
\end{equation}
where $|x|_1$ denotes the $L^1$-norm of an array $x$. Intuitively, the increment gives a measure of how well the solution has converged after $n$ iterations. Finally, we set
\begin{equation}
    \lambda_{n+1} = \begin{cases}
\lambda_n\, & \text{if}\quad\text{increment}^{(n+1)} \leq \text{increment}^{(n)}\\
\lambda_n/2\,  & \text{if}\quad\text{increment}^{(n+1)} > \text{increment}^{(n)}
\end{cases}\ .
\end{equation}
The role of the mixing parameter $\lambda_n$ is to stabilize the numerical procedure and was already emphasized in~\cite{Maldacena:2016hyu}.
\smallskip

\section{Singularities of the two-point function}
\label{S:Singularities of the two-point function}

We now have the basic framework for analyzing the model. In this section, we present our results for the thermal two-point functions and discuss some of its properties. In particular, we will focus on the singularities in the complex $\tau$-plane for various values of the temperature.

\subsection{Two-point function: double expansion and numerics}

\begin{figure}[htbp!]
  \begin{subfigure}[b]{0.5\textwidth}
    \centering
    \includegraphics[width=0.9\linewidth]{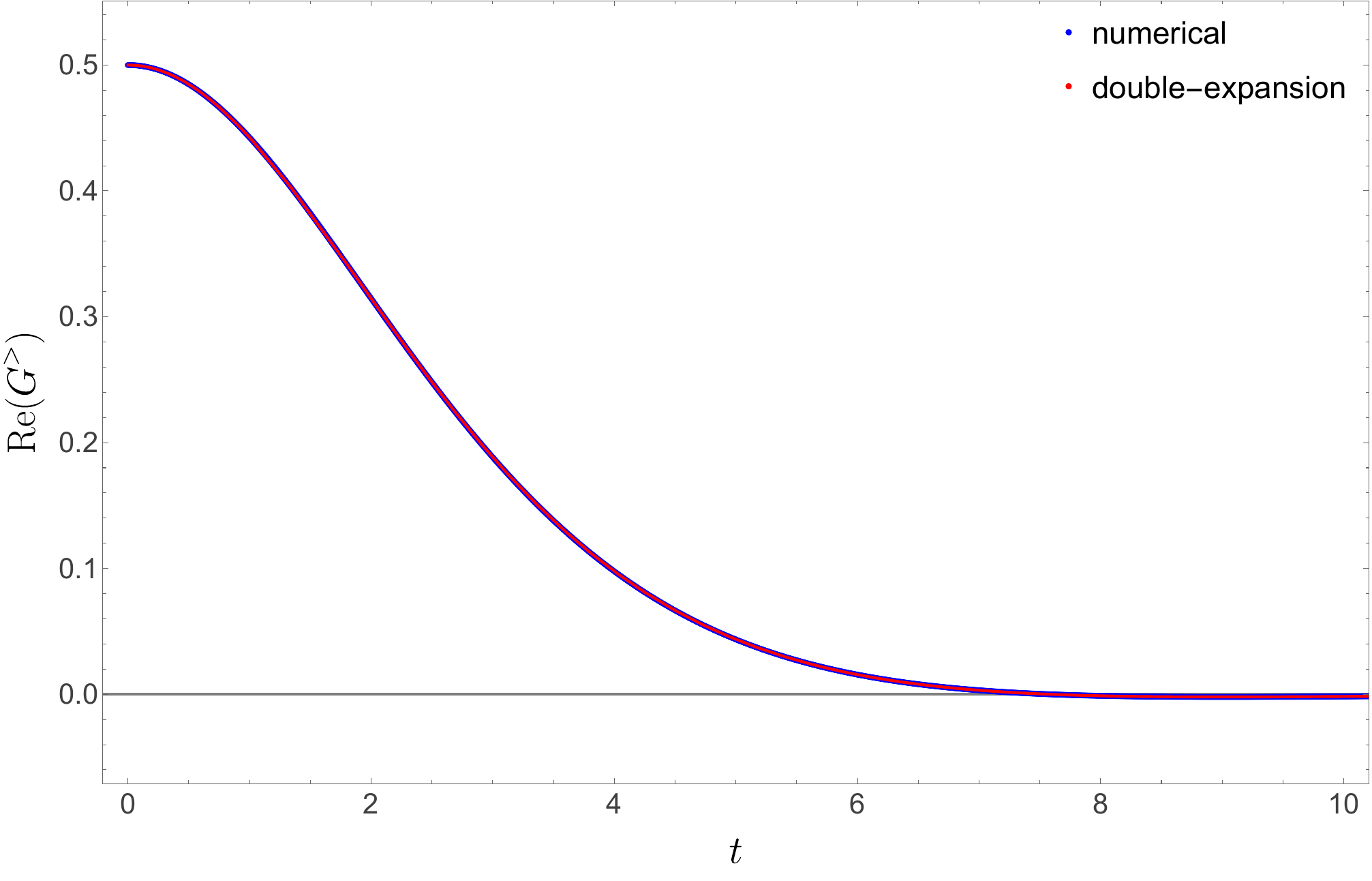}
    \caption{$\mathrm{Re}\,G^>(t)$}
    \label{betaJ-1/2-real}
\end{subfigure}
\hfill
 \begin{subfigure}[b]{0.5\textwidth}
    \centering
    \includegraphics[width=0.9\linewidth]{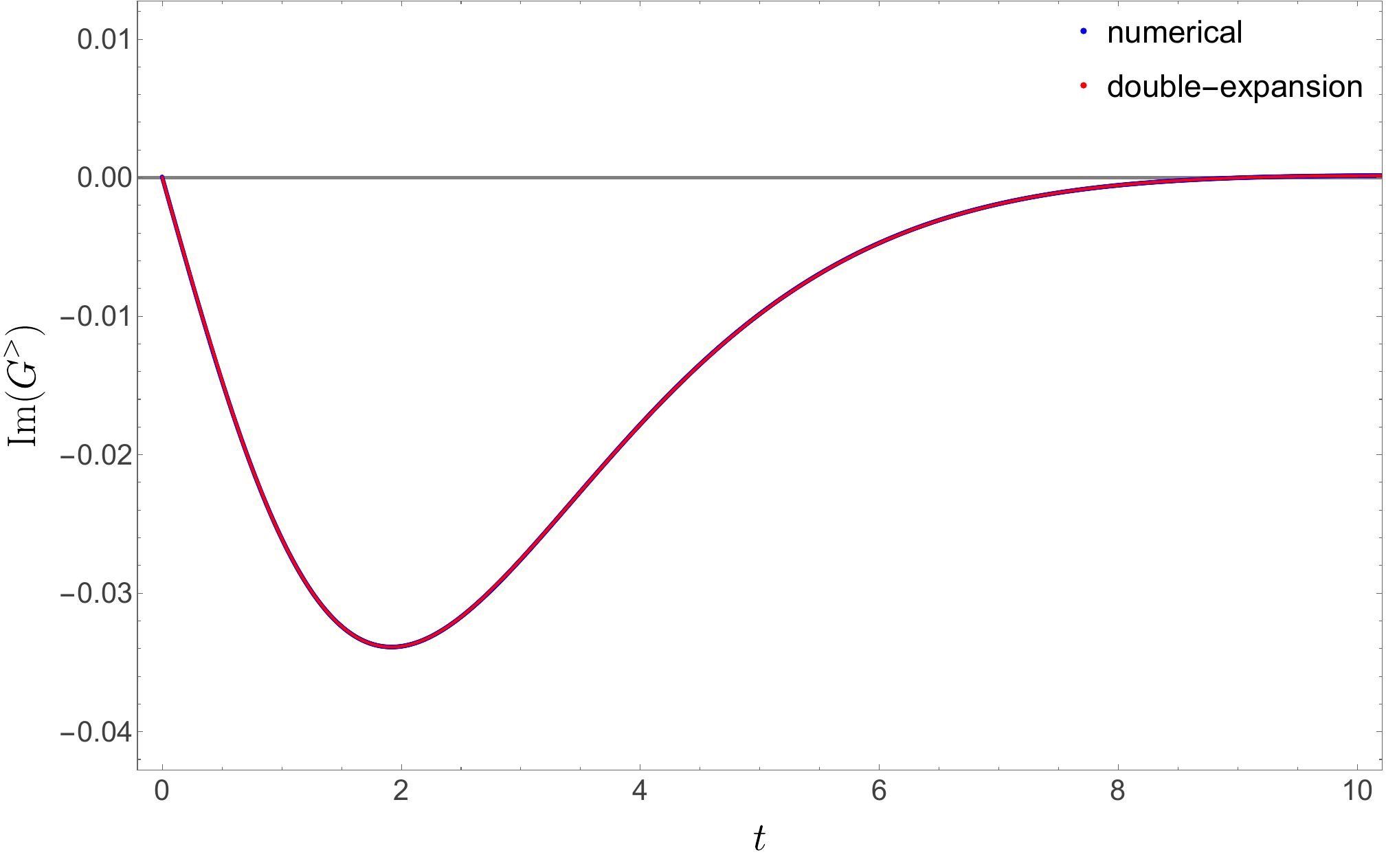}
    \caption{$\mathrm{Im}\,G^>(t)$}
    \label{betaJ-1/2-imaginary}
    \end{subfigure}
\caption{Comparison of the numerical real-time Schwinger--Dyson solution (blue) with the double-expanded Pad\'e approximation (red) for $G^>(t)$. Parameters: $q=4$ and $\beta=1/2$.}
\label{compare-betaJ-1/2}
\end{figure}

\begin{figure}[h]
    \begin{subfigure}[b]{0.5\textwidth}
    \centering
    \includegraphics[width=0.8\linewidth]{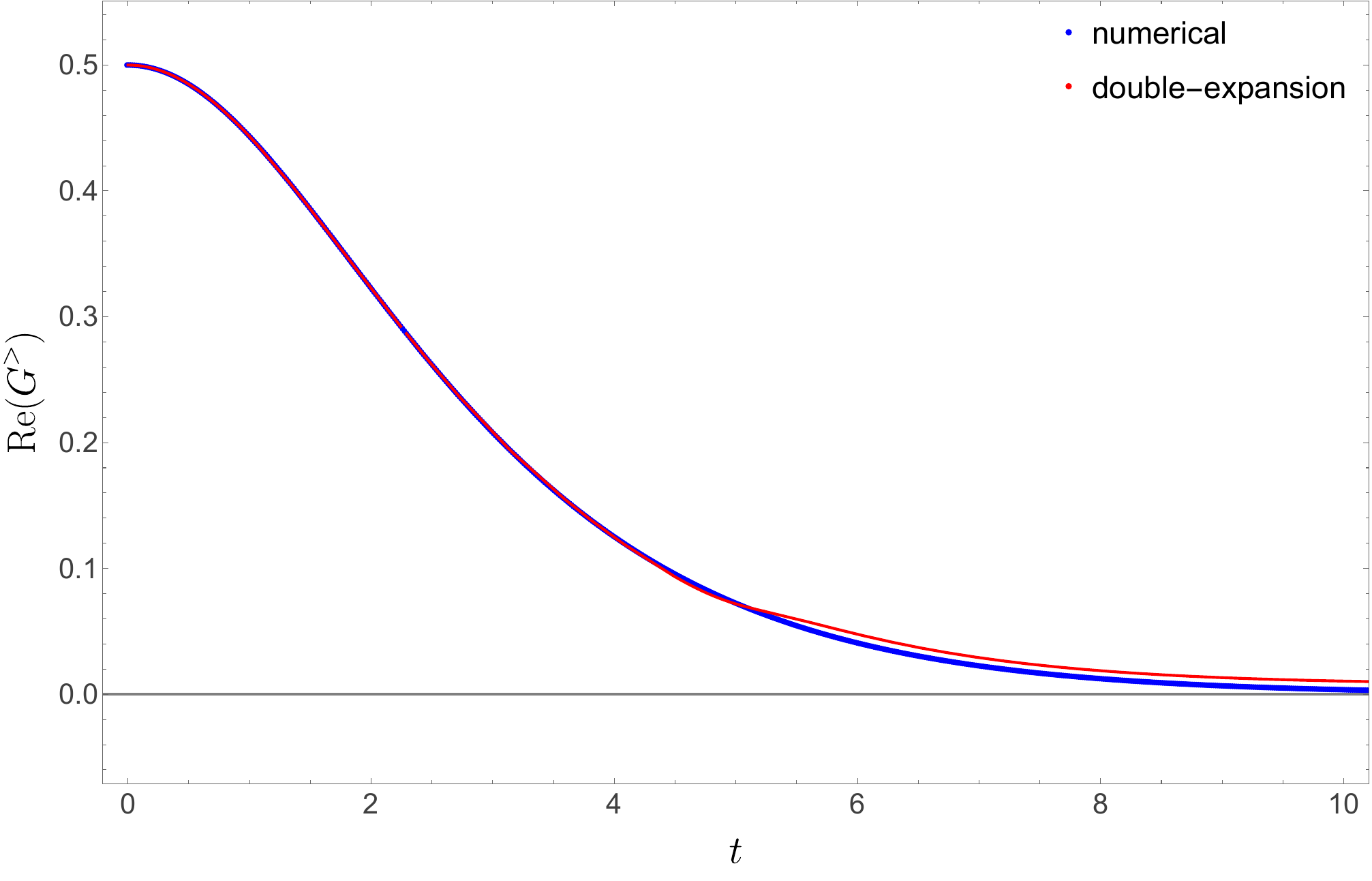}
    \caption{$\mathrm{Re}\,G^>(t)$}
    \label{betaJ-2-real}
\end{subfigure}
\hfill
 \begin{subfigure}[b]{0.5\textwidth}
    \centering
\includegraphics[width=0.8\linewidth]{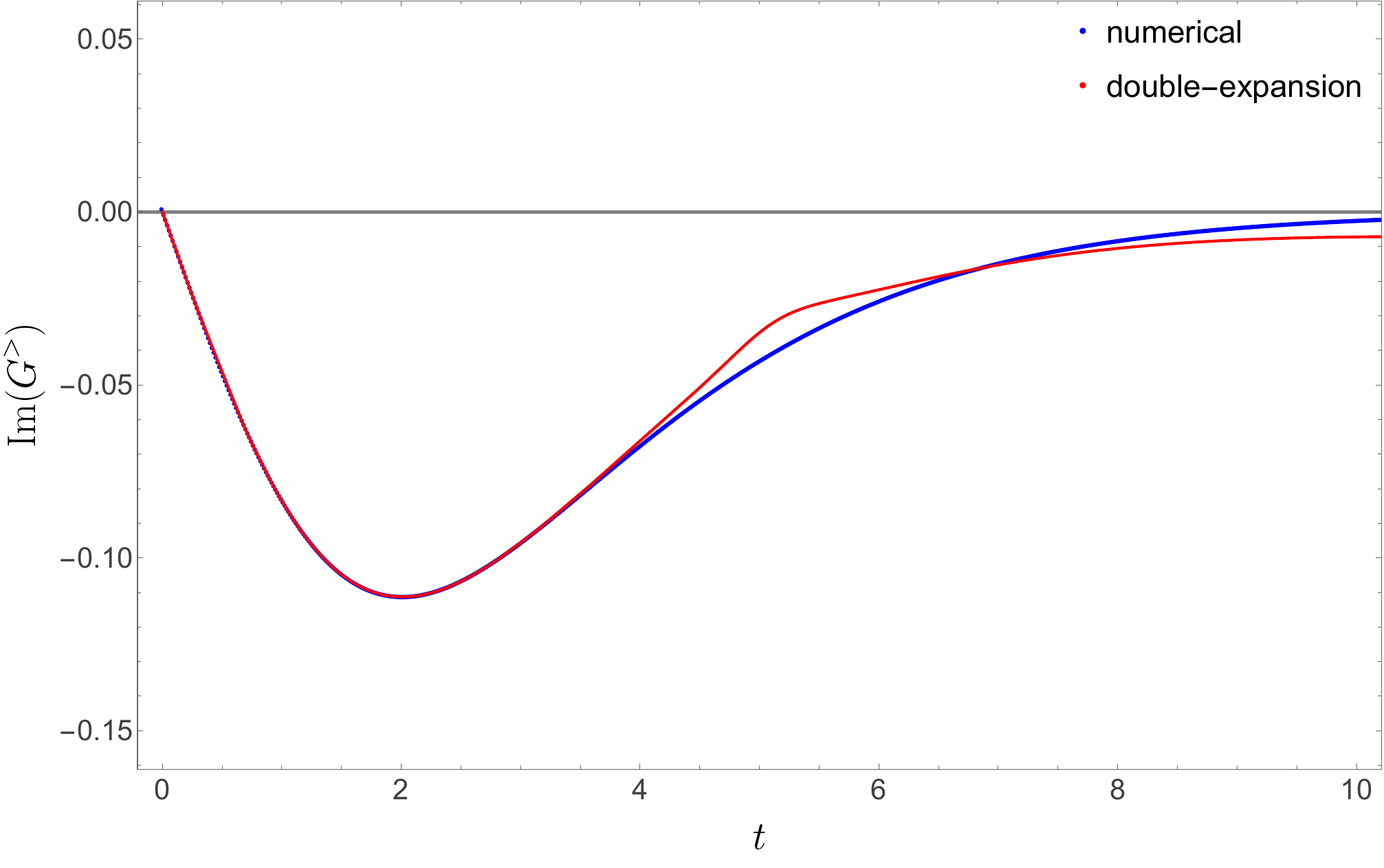}
\caption{$\mathrm{Im}\,G^>(t)$}
     \label{betaJ-2-imaginary}
    \end{subfigure}
    \caption{Comparison of the numerical real-time Schwinger--Dyson solution (blue) with the double-expanded Pad\'e approximation (red) for $\mathrm{Re}\,G^>(t)$. Parameters: $q=4$ and $\beta=2$.}
    \label{compare-betaJ-2}
\end{figure}

Both methods that we have presented in the previous section converge to give the thermal two-point function of the SYK model. As a first sanity check, we verify that the two-point functions obtained numerically and via double-expansion agree with one another. This is indeed the case. In Figs.~\ref{compare-betaJ-1/2}-\ref{compare-betaJ-2} we compare approximations to the Wightman function $G^>(t)$ obtained by the two methods for some fixed values of $\beta$. For the smaller value $\beta=1/2$, the two curves lie on top of one another. For the larger value $\beta=2$, we observe excellent agreement for Lorentzian times $t\lesssim5$ and some deviations at late times.

\subsection{Location of singularities}\label{Singularity Location}

We begin with analyzing the singularities of the two-point function in the complex $\tau$-plane for a fixed value of $\beta$. Then, we will increase $\beta$ from zero, where the singularities have been obtained previously in~\cite{Dodelson:2024atp}, towards infinity.

In the double-expansion method, singularities naturally arise through the use of Pad\'e approximation. Namely, by virtue of the construction, for any finite order of truncation, the resulting approximation of the two-point function is a rational function of $\tau$ and thus has poles at the roots of the denominator polynomial. In this approach, the true singularities of the correlator are obtained as poles of the Pad\'e approximant that converge under increase of the truncation order. In the numerical approach, the two-point function is determined for real values of either Euclidean or Lorentzian time. Therefore, generic singularities, which are complex, cannot be directly observed from the plot of the numerical solution. Instead, the latter is approximated by a rational function using the so-called AAA algorithm, \cite{AntoulasAnderson1986,Nakatsukasa_2018}. The AAA algorithm is more stable than interpolation followed by a rational approximation and suffers less from numerical artifacts such as Froissart doublets.
\smallskip

\begin{figure}[hbtp!]
\begin{subfigure}[b]{0.5\textwidth}
\centering
    \includegraphics[width=0.9\linewidth]{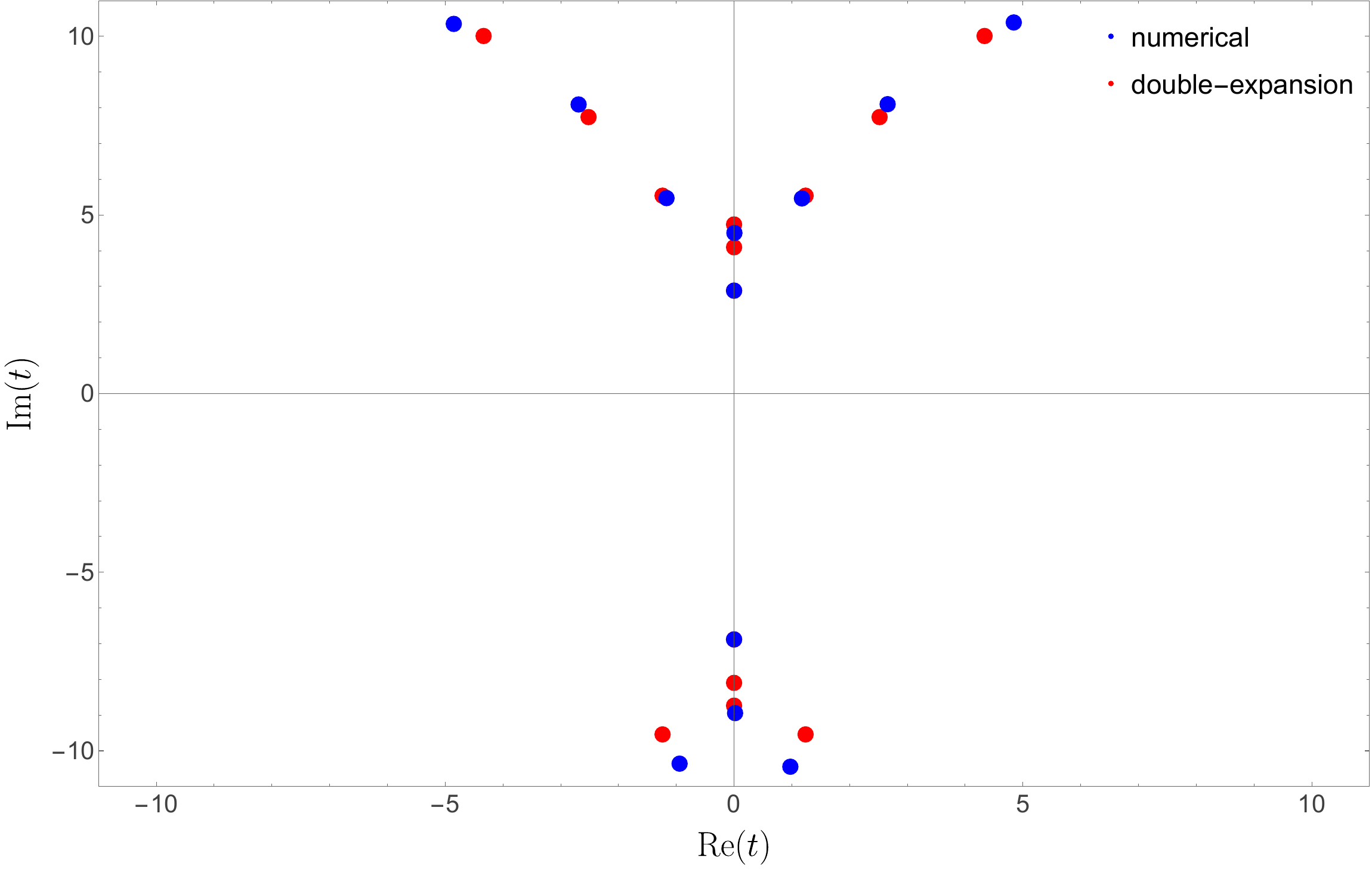}
    \caption{Parameters: $q=4$ and $\beta=4$}
      \label{compare-poles-betaJ-4}
\end{subfigure}
\begin{subfigure}[b]{0.5\textwidth}
\centering    
    \includegraphics[width=0.9\linewidth]{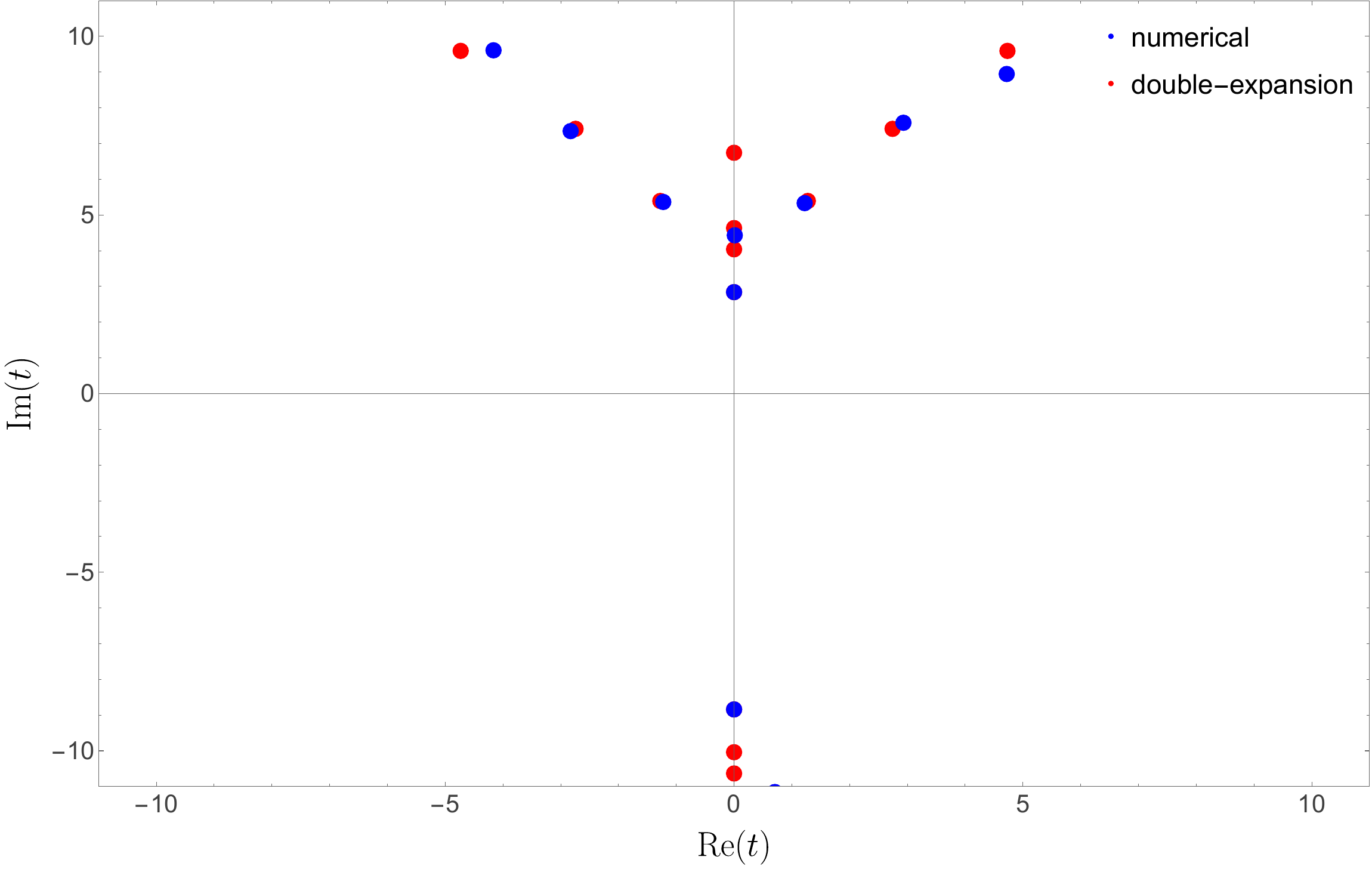}
    \caption{Parameters: $q=4$ and $\beta=6$}
        \label{compare-poles-betaJ-6}
\end{subfigure}
\caption{Comparison of poles in the two-point function given by numerical solution (blue) and the Pad\'e approximated double-expansion (red). }
    \label{compare-poles-betaJ-4and6}
\end{figure}

As in the previous subsection, we can plot the singularities of the two-point function arising from either method and compare the results. A comparison of poles for $\beta=4$ and $\beta=6$ are shown in Figs.~\ref{compare-poles-betaJ-4} and~\ref{compare-poles-betaJ-6}, respectively. For the leading singularity, i.e., the lowest one on the positive imaginary axis, only one point is visible as the results of the two approaches lie on top of one another. While some other lowest-lying poles may clearly be identified between the two approaches, the higher poles are increasingly difficult to stabilize. In the remainder of this section, we will focus on the leading and first two subleading poles and track their location as $\beta$ is varied from zero to infinity.

\subsection{Temperature dependence of low-lying singularities}

By leading and first two subleading poles, we refer to the following three poles in upper half $t$-plane. The leading pole is the one on the imaginary axis with the lowest positive imaginary value. The subleading poles are the poles off the imaginary axis with lowest positive imaginary values, see Fig.~\ref{compare-poles-betaJ-4and6}. Since the subleading poles lie symmetrically around the imaginary axis, it suffices to focus on one of them, e.g., the one with positive real part. 
\smallskip

\begin{figure}[htp!]
    \centering
    \includegraphics[width=0.68\linewidth]{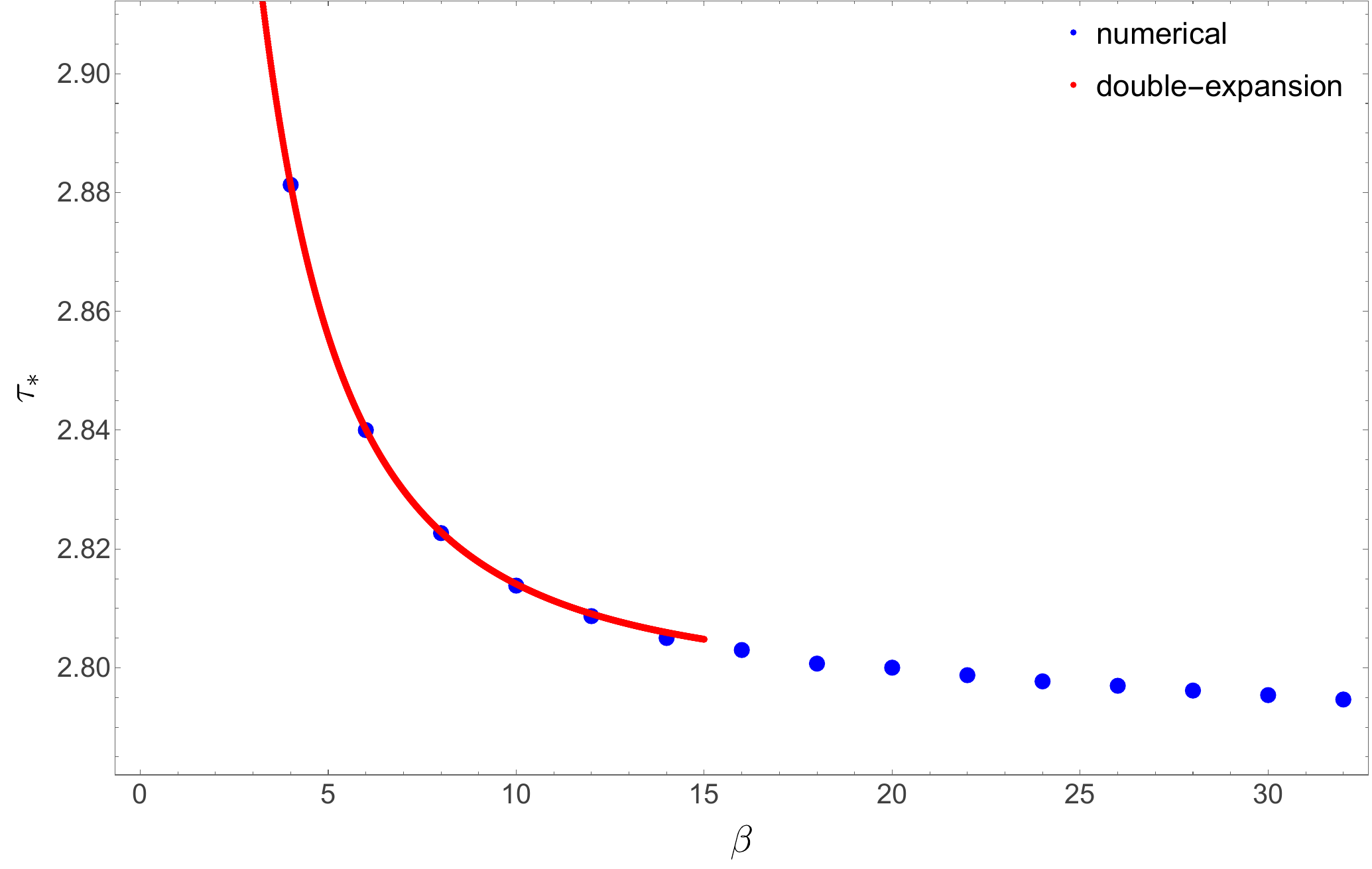}
    \caption{Dependence of the imaginary part $\tau_*$ of the leading pole on $\beta$.}
    \label{leading-pole-dependence}
\end{figure}

\begin{figure}[hbt!]
    \centering
    \begin{subfigure}[b]{0.49\textwidth}
        \includegraphics[width=0.9\linewidth]{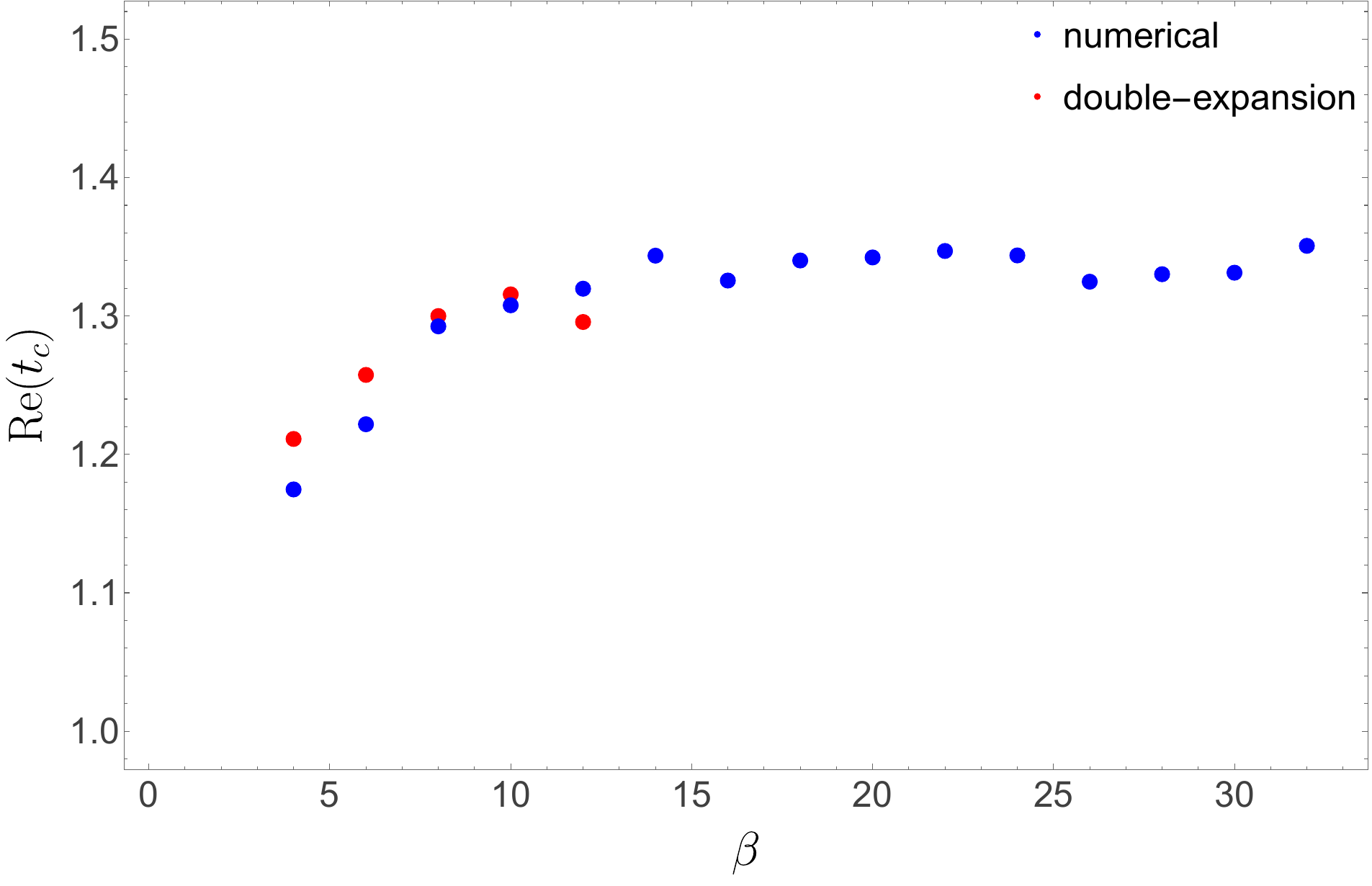}
        \caption{$\Re(t_c)$}
    \end{subfigure}
    \hfill
    \begin{subfigure}[b]{0.49\textwidth}
        \includegraphics[width=0.9\linewidth]{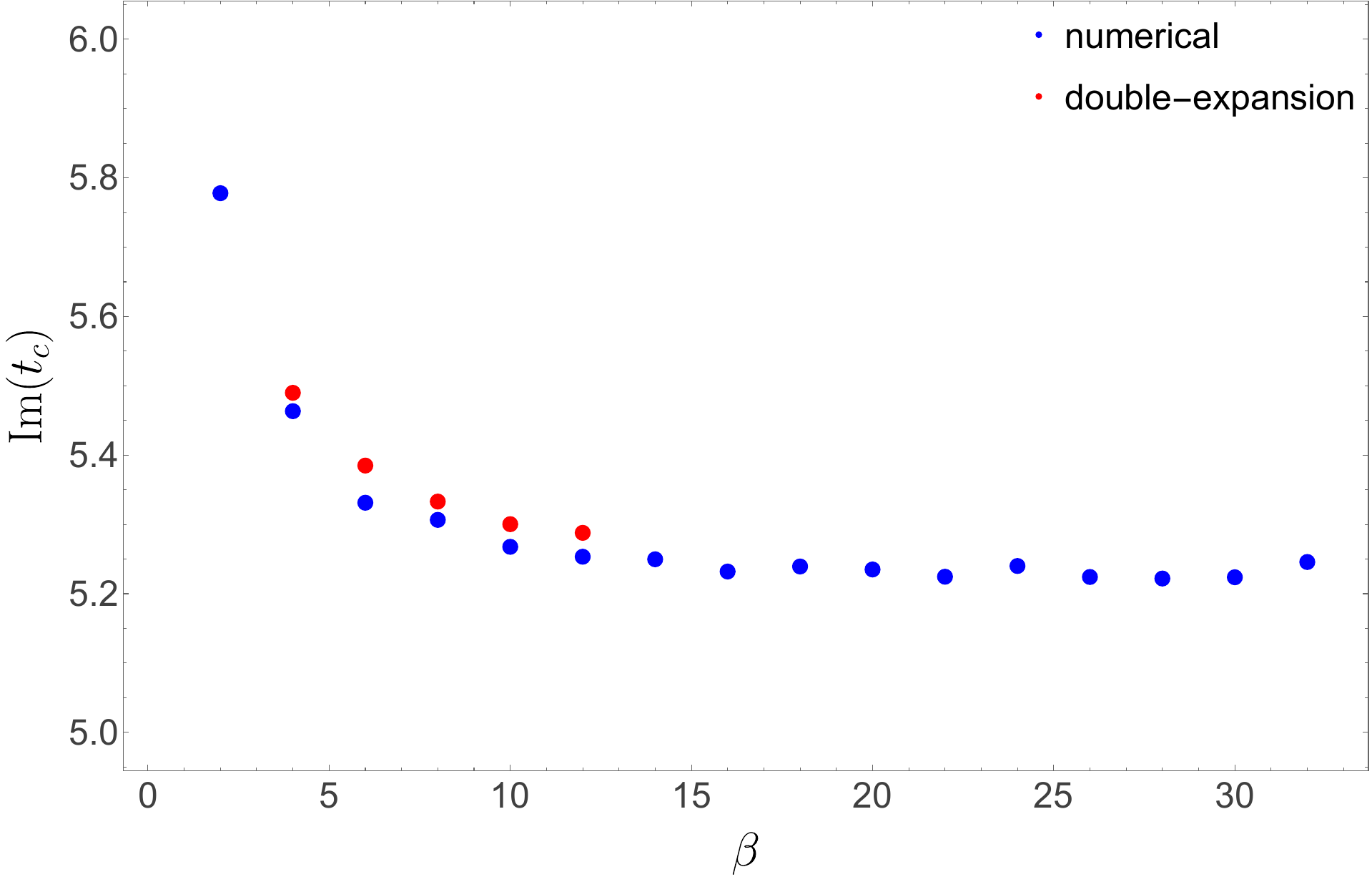}
        \caption{$\Im(t_c)$}
    \end{subfigure}
    \caption{Dependence of the first subleading singularity on $\beta$. As in the earlier plots, the double-expansion results are plotted for comparison for lower values of $\beta$.}
    \label{bouncing-singularity-plot}
\end{figure}

We shall denote the location of the leading pole by $i\tau_*$, so that $\tau_*>0$. It turns out that $\tau_*$ only mildly varies with $\beta$ and seems to approach a constant value $\sim2.8$ as $\beta\to\infty$, although the numerical approach cannot be pushed to values of $\beta$ much higher than 30. The dependence of $\tau_*$ on $\beta$ is shown on Figure \ref{leading-pole-dependence}. Similarly, the first subleading pole, which we denote by $t_c=t_c(\beta)$, appears to converge as $\beta\to\infty$. Its real and imaginary parts are shown on Figure \ref{bouncing-singularity-plot}. In both of these figures, we display results obtained both by the double expansion and by the numerical approach. The results between them agree well for high temperatures $\beta\lesssim12$, for which the Pad\'e approximated double expansion has converged. Above this value of $\beta$, we only use the numerical approach.
\smallskip

As discussed in~\cite{Parker:2018yvk} and the next subsection, the leading singularity is precisely the decay constant of the spectral density $\rho(\omega) \sim e^{-\tau_* |\omega|}$ at large frequency. In Appendix C of \cite{Gu:2021xaj}, the Schwinger-Dyson equations were solved numerically and the best exponential fit of $\rho(\omega)$ was performed. They similarly found that $\tau_*$ approaches a constant value $\sim 2.8/J$ at low temperatures, and their plot of $\tau_* J$ vs $\beta J$ shows good agreement with our Figure \ref{leading-pole-dependence}. 

\subsection{Structure of the leading singularity}

In this subsection, we give an analytic argument for the nature of the nearest singularity of the zero-temperature two-point function. While the argument cannot be used to determine the actual numerical value of the singularity location, it fixes the leading-order power and the residue of the nearest singularity expected from the Schwinger--Dyson solution. We first keep the interaction order $q$ general and then specialize to the large-$q$ and $q=4$ limits.
\smallskip

For positive Euclidean time, the zero-temperature correlator admits the spectral representation,
\begin{equation}
  G(\tau)=\int_0^\infty \frac{d\omega}{2\pi}\,\rho(\omega)e^{-\omega \tau}\,,
  \qquad \tau>0\ .
\end{equation}
The physical zero-temperature Euclidean correlator is holomorphic for $\operatorname{Re}\tau>0$. We assume that the nearest singularity of its analytic continuation lies on the negative Euclidean-time axis at $\tau=-\tau_*<0$. This is equivalent to assuming that the large-frequency behavior of the positive-frequency spectral density has the WKB form\footnote{This is also equivalent to the universal operator growth hypothesis, \cite{Parker:2018yvk}.}
\begin{equation}\label{WKB}
  \rho(\omega)\sim A\,\omega^\alpha e^{-\tau_*\omega}\,,
  \qquad \omega\to+\infty\,,
\end{equation}
for some constants $\alpha$ and $A$. The exponent $\tau_*$ controls the distance to the nearest singularity, while the power $\alpha$ controls its local type. 

At large real frequency, the retarded correlator $G_R(\omega)$ is generated, to leading order, by the self-energy,
\begin{equation}
  G_R(\omega) =
  \frac{1}{-i\omega-i\Sigma_R(\omega)} \simeq \frac{i}{\omega} - \frac{i\Sigma_R(\omega)}{\omega^2} + \cdots\ .
\end{equation}
Therefore, the spectral density behaves as
\begin{equation}
  \rho(\omega) = 2\operatorname{Re}G_R(\omega) \sim \frac{2\operatorname{Im} \Sigma_R(\omega)}{\omega^2} = \frac{\rho_\Sigma(\omega)}{\omega^2}\,,
\end{equation}
where the self-energy spectral density $\rho_\Sigma$ appearing in the last line is defined by
\begin{equation}\label{self-energy-density}
    \Sigma(\tau) = \int_0^\infty \frac{d\omega}{2\pi}\, \rho_\Sigma(\omega)e^{-\omega\tau}\ .
\end{equation}
In order to determine the large-$\omega$ behavior of $\rho(\omega)$, we use the zero-temperature spectral representation of the two-point function together with $\Sigma(\tau)=G(\tau)^{q-1}$ to write
\begin{equation}
    \Sigma(\tau)=  \int_0^\infty \prod_{a=1}^{q-1} \frac{d\omega_a}{2\pi}\,\rho(\omega_a)\, e^{-\tau\sum_a\omega_a}\ .
\end{equation}
Comparing the last equation with \eqref{self-energy-density}, we read off
\begin{equation}
    \rho_\Sigma(\omega) = \int_0^\infty \frac{\prod_{a=1}^{q-1}d\omega_a}{(2\pi)^{q-2}}\, \prod_{a=1}^{q-1}\rho(\omega_a)\,
    \delta\!\left(\omega-\sum_{a=1}^{q-1}\omega_a\right)\ .
\end{equation}
By substituting the WKB form \eqref{WKB} for each $\rho(\omega_a)$ and performing the integral, we find
\begin{equation}
  \rho_\Sigma(\omega) \sim
  \frac{A^{q-1}}{(2\pi)^{q-2}}
  \frac{\Gamma(\alpha+1)^{q-1}}{\Gamma((q-1)(\alpha+1))}
  \omega^{(q-1)\alpha+q-2} e^{-\tau_*\omega}\ .
\end{equation}
Consequently, the spectral density behaves as
\begin{equation}\label{reproducing-WKB}
  \rho(\omega) \sim \frac{\rho_\Sigma(\omega)}{\omega^2}
  \sim
  \frac{A^{q-1}}{(2\pi)^{q-2}}
  \frac{\Gamma(\alpha+1)^{q-1}}{\Gamma((q-1)(\alpha+1))}
  \omega^{(q-1)\alpha+q-4} e^{-\tau_*\omega}\ .
\end{equation}
The WKB ansatz \eqref{WKB} can only be self-consistent if it is reproduced by \eqref{reproducing-WKB}. Matching the power of $\omega$ and the coefficient fixes
\begin{equation}\label{pow and amp}
    \alpha = -\frac{q-4}{q-2}\,, \qquad
    A = 2\pi\, \Gamma \left(2+\frac{2}{q-2}\right)^{\frac{1}{q-2}} \Gamma
   \left(\frac{2}{q-2}\right)^{\frac{1}{2-q}-1}\ .
\end{equation}
The structure of the leading singularity of the correlator is then
\begin{equation}\label{eq:singularity-structure-arbitrary-q}
G(\tau) \sim A\int^\infty \frac{d\omega}{2\pi}\, \omega^\alpha  e^{-\omega(\tau+\tau_*)}  \sim \frac{A \ \Gamma(1+\alpha)}{2\pi(\tau+\tau_*)^{1+\alpha}}= \left( \frac{2q}{(q-2)^2} \right)^{\frac{1}{q-2}}\frac{1}{(\tau+\tau_*)^{\frac{2}{q-2}}}\ .
\end{equation}
This is the same singularity structure observed numerically at infinite temperature in \cite{Dodelson:2025jff}, but here we provide an analytic prediction for the residue.
\smallskip

The large-$q$ limit provides a useful check on the normalization. Restoring the $J$-dependence in \eqref{eq:singularity-structure-arbitrary-q}, the leading singular behavior is
\begin{equation}
G(\tau)\sim
\left( \frac{2q}{J^2(q-2)^2} \right)^{\frac{1}{q-2}}
\frac{1}{(\tau+\tau_*)^{\frac{2}{q-2}}}\ .
\end{equation}
This can be compared with the exact zero-temperature large-$q$, fixed $\mathcal{J} = \tfrac{\sqrt{q}}{2^{(q-1)/2}} J$, limit of the correlator obtained in \cite{Maldacena:2016hyu},
\begin{equation}
    G(\tau) = \frac{1}{2} - \frac{1}{q}\log(\mathcal{J} \tau +1) + \mathcal{O}\left(\tfrac{1}{q^2}\right) \ .
\end{equation}
Taking the same large-$q$ limit of the general expression gives, assuming that $\tau_*$ depends sufficiently mildly on $q$,
\begin{equation}
    \left( \frac{2q}{J^2(q-2)^2} \right)^{\frac{1}{q-2}}\frac{1}{(\tau+\tau_*)^{\frac{2}{q-2}}} \underset{q\to\infty}{=} \frac{1}{2} - \frac{1}{q}\log(\mathcal{J} (\tau +\tau_*)) + \mathcal{O}\left(\tfrac{1}{q^2}\right)\ .
\end{equation}
We see that equation \eqref{eq:singularity-structure-arbitrary-q} correctly reproduces the residue of the $\tau = -\tau_*$ branch point at leading nontrivial order in large $q$. We also note from the exact solution that $\tau_* = 1/\mathcal{J}$.
\smallskip

For $q=4$, \eqref{pow and amp} gives
\begin{equation}
    \alpha=0\,, \qquad A = 2\sqrt{2}\pi\ .
\end{equation}
Going back to the Euclidean correlator, the leading singular behavior is therefore
\begin{equation}\label{WKB-singularity-predicition}
  G(\tau) \sim A\int^\infty \frac{d\omega}{2\pi}\, e^{-\omega(\tau+\tau_*)} = \frac{\sqrt{2}}{\tau+\tau_*}\ .
\end{equation}
This last step should be read as determining the leading local behavior implied by the leading WKB tail, rather than the full analytic structure near the singularity. In the infinite-temperature analysis of \cite{Dodelson:2025jff}, the corresponding $q=4$ singularities were found numerically to be branch points with local behavior of the form
\[
    G(\tau)\sim \frac{a}{\tau+\tau_*}+b\log(\tau+\tau_*)+\cdots\,,
\]
after translating to the present notation. Thus, the pole term is still the most singular term, while the logarithm signals that the singularity is not an isolated pole. Such logarithmic terms would naturally arise from subleading corrections to the large-$\omega$ spectral density, for example terms proportional to $e^{-\tau_*\omega}/\omega$. The simple scaling argument above only fixes the leading pole coefficient.
\smallskip

\begin{figure}[h]
    \centering
        \includegraphics[width=0.68\linewidth]{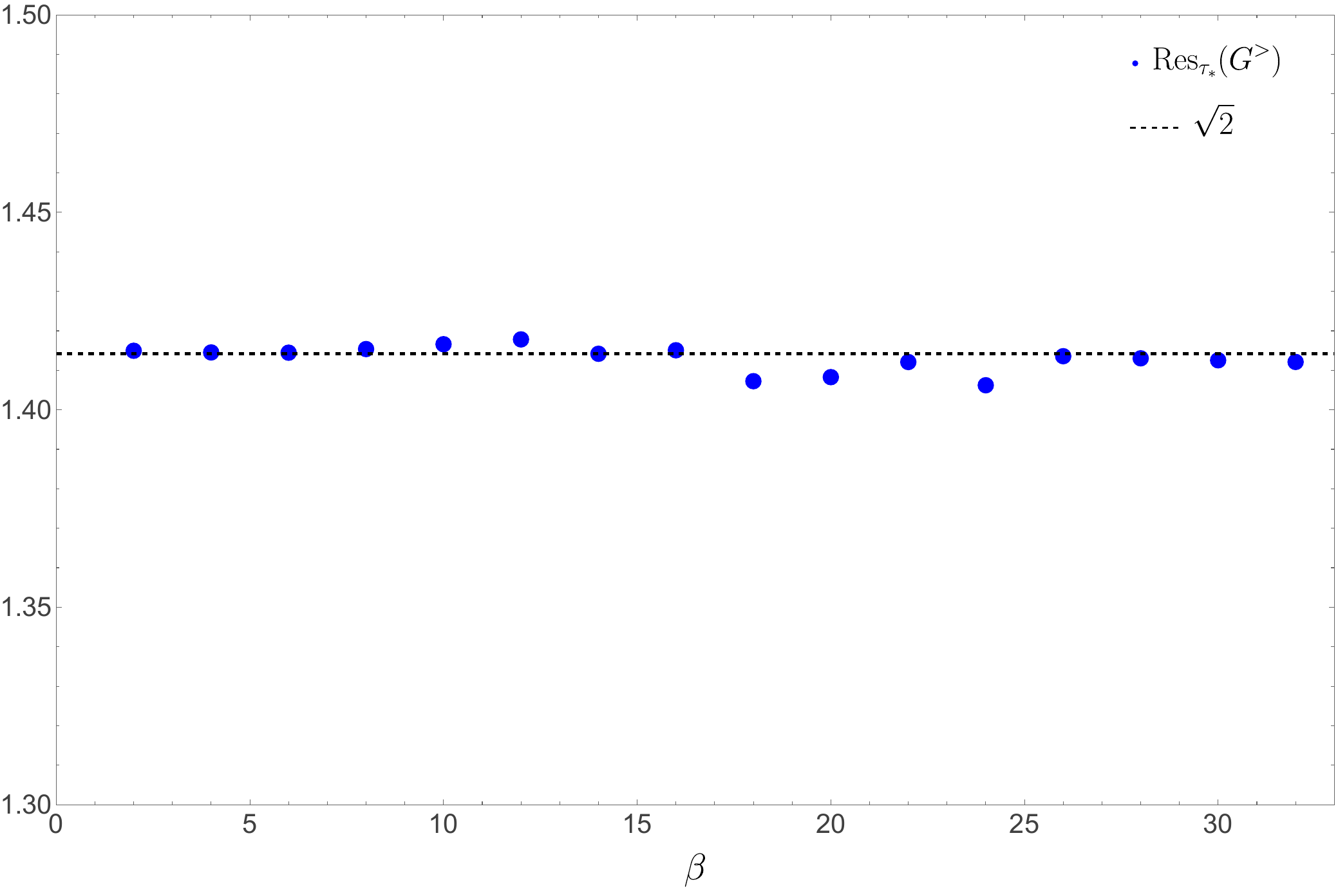}
    \caption{Numerical estimate of the residue of the leading pole. Parameters: $q=4$.}
    \label{leading-residue-plot}
\end{figure}

The above argument goes through even at finite temperature. In this case, the Euclidean correlator has a different spectral representation, namely
\begin{equation}
  G(\tau) = \int_{-\infty}^\infty \frac{d\omega}{2\pi}\, \frac{\rho(\omega)e^{-\omega\tau}}
  {1+e^{-\beta\omega}}\,, \qquad 0<\tau<\beta\ .
\end{equation}
Nevertheless, due to the high-frequency behavior of the Boltzmann factor,
\begin{equation}
  \frac{1}{1+e^{-\beta\omega}} \longrightarrow 1\,,\qquad \omega\to+\infty\,,
\end{equation}
the finite-temperature generalization of the argument gives the same $q$-dependent power and amplitude found in \eqref{pow and amp}. However, at finite temperature, the two-point function has additional singularities - the KMS reflections of $\tau_*$. In particular, there is a singularity at
\begin{equation}
    \tau = \beta + \tau_*\,,
\end{equation}
which is the leading singularity in the lower half-plane on Figure \ref{compare-poles-betaJ-4and6}. While the behavior of the correlator \eqref{WKB} and its consequences in position space were not derived from first principles, they are supported by our numerical results. Namely, for $q=4$ we observe that the residue of the leading pole is numerically very close to $\sqrt{2}$, consistently with \eqref{WKB-singularity-predicition} and as illustrated in~Fig.~\ref{leading-residue-plot}.

\section{Discussion}
\label{S:Discussion}

In this paper, we analyzed the finite temperature two-point function of the fundamental fermion in the large-$N$ SYK model. Focusing on the little explored UV regime, we solved the Schwinger-Dyson equations by two complementary methods - via double-expansion and numerically. The two methods show excellent agreement and reveal singularities of the two-point function in the complex $\tau$-plane. We determined the temperature dependence of the lowest lying singularities, extending previous work for infinite temperature~\cite{Dodelson:2025jff}. In the remainder of this section, we discuss some applications of the methods developed herein and potential holographic interpretations of the results.

\paragraph{Thermodynamic applications:} While in this work we have focused on the analytic structure of the two-point function, the same Pad\'e-resummed double expansion can also be used to study thermodynamic observables. In particular, it gives a complementary way of approaching the low-temperature expansion of the SYK energy considered in
\cite{Cruz:2022uic}. The key observation is that, at large $N$, the energy may be written in terms of the two-point function,
\begin{equation}
    E = -\frac{J^2}{q}\int_0^\beta d\tau\, G(\tau)^q = -\frac{1}{q}\,\partial_\tau G(\tau)\big|_{\tau\to0^+}\ .
\end{equation}
The analysis of \cite{Cruz:2022uic} used high-precision numerical solutions of the finite-temperature Schwinger-Dyson equations to extract the low temperature asymptotic expansion of the dimensionless energy
\begin{equation}\label{low energy expansion}
    \epsilon \equiv \frac{E}{J} = c_0+\frac{c_2}{(\beta J)^2}+\frac{c_3}{(\beta J)^3}+\cdots\,,
\end{equation}
and to determine which integer and non-integer powers are present. The Pad\'e-resummed double expansion provides an independent way to extract the same coefficients. To this end, we evaluate $\approx100$ iterations of the double-expansion algorithm and study the linear term in $\tau$ (see e.g. \eqref{G tau beta}). The latter is a series in $\beta$. After resummation of this series, the energy can be evaluated at moderately high values of $\beta J$. One may then fit the resulting values of $E(\beta)$ to the low-temperature form \eqref{low energy expansion}. The results are presented in Table \ref{tab:energy-coefficients-comparison}.
\begin{table}[h]
    \centering
    \begin{tabular}{c|c|c}
        \hline
        coefficient & Pad\'e-resummed double expansion & \cite{Cruz:2022uic} \\
        \hline
        $c_0$ & $-0.0406303$ & $-0.0406302697583$ \\
        $c_2$ & $\phantom{-}0.19801$ & $\phantom{-}0.198008397003$ \\
        $c_3$ & $-0.419$ & $-0.419469896737$ \\
        $c_4$ & $\phantom{-}0.67$ & $\phantom{-}0.664982720600$ \\
        $c_5$ & $-2.6$ & $-2.57685914760$  \\
        \hline
    \end{tabular}
    \caption{Comparison of the coefficients in the low-temperature expansion of the energy, \eqref{low energy expansion},
    extracted from the Pad\'e-resummed double expansion and from the numerical analysis of \cite{Cruz:2022uic}.}
    \label{tab:energy-coefficients-comparison}
\end{table}

\paragraph{Kinematic space interpretation:}

For the $q=4$ model at zero temperature, we observed the leading complex singularity of the analytically continued Green's function  located at
\begin{equation}
t_* \simeq \frac{2.8 i}{J}\ .
\end{equation}
This singularity is a microscopic feature of the exact SYK Green's function, and it is therefore absent in the purely conformal approximation. Nevertheless, it can be seen as a direct consequence for the bilocal Green's function interpreted in kinematic space language~\cite{Maldacena:2016hyu}.  
\smallskip

To see this, consider the four-point function in the large-$N$ expansion
\begin{equation}
\frac{1}{N^2} \sum_{i,j=1}^N
\left\langle T\left(\psi_i(\tau_1)\psi_i(\tau_2)\psi_j(\tau_3)\psi_j(\tau_4)
\right)\right\rangle=
G(\tau_{12})G(\tau_{34})+\frac{1}{N}F(\tau_1,\tau_2,\tau_3,\tau_4)+\cdots\ .
\end{equation}
The $O(1/N)$ connected piece $F(\tau_1,\dots,\tau_4)$ is obtained with the help of the ladder kernel, namely
\begin{equation}
F = \frac{1}{1-\mathcal{K}}F_0\,, \qquad F_0(\tau_1,\tau_2,\tau_3,\tau_4)
= G(\tau_{14})G(\tau_{23})-G(\tau_{13})G(\tau_{24})\ .
\end{equation}
Here, $\mathcal{K}$ is the integral operator
\begin{equation}
    (\mathcal{K}f)(\tau_1,\dots,\tau_4) = \int d\tau d\tau'\, K(\tau_1,\tau_2;\tau,\tau') f(\tau,\tau',\tau_3,\tau_4)\,,
\end{equation}
and $K$ is the ladder kernel
\begin{equation}
K(\tau_1,\tau_2;\tau_3,\tau_4) = -J^2(q-1)G(\tau_{13})G(\tau_{24})G(\tau_{34})^{q-2}\ .
\end{equation}
It is through the ladder kernel that the relation to geometry starts to emerge. To see this, one introduces coordinates
\begin{equation}\label{eq:bilocal-coordinates-discussion}
T=\frac{t_1+t_2}{2}\,,\quad z=\frac{t_1-t_2}{2}\,, \qquad T'=\frac{t_3+t_4}{2}\,,
\quad
z'=\frac{t_3-t_4}{2}\ .
\end{equation}
This structure is especially transparent in the large-$q$ limit, where the ladder kernel can be evaluated explicitly:
\begin{equation}
K(\tau_1,\tau_2;\tau_3,\tau_4)
= -\frac{\mathcal J^2}{2}\operatorname{sgn}(\tau_{13})
\operatorname{sgn}(\tau_{24})e^{g(\tau_{34})} \ .
\end{equation}
The factor $e^{g(\tau_{34})}$ is the large-$q$ limit of the rung factor $G(\tau_{34})^{q-2}$. More precisely, in the standard large-$q$ normalization,
\begin{equation}
\left(2G(\tau)\right)^{q-2} = e^{g(\tau)}\left(1+\mathcal O(q^{-1})\right)\ .
\end{equation}
Furthermore, at zero temperature we have
\begin{equation}
e^{g(\tau)} =\frac{1}{(1+\mathcal J\tau)^2}\,,
\end{equation}
for positive Euclidean time before analytic continuation. Thus, $e^{g(\tau)}$ has a double pole at $\tau=-\mathcal J^{-1}$.

This form gives the starting point for the kinematic-space discussion~\cite[Appendix I]{Maldacena:2016hyu}.  After passing to the bilocal coordinates in \eqref{eq:bilocal-coordinates-discussion} and shifting the radial variable by the microscopic scale $\mathcal J^{-1}$, the large-$q$ kernel can be written in terms of the regulated AdS$_2$ propagator. In this description, the same short-time scale that appears in the double pole of
$e^{g(\tau)}$ becomes a radial cutoff in kinematic space. This observation provides the point of departure for the finite-$q$ discussion below.

At $q=4$, a closely analogous mechanism is visible directly in the finite-$q$ ladder kernel. The latter contains the factor $G(t_{34})^2 = G(2z')^2$. Thus, the singularity of the two-point function appears in the complexified
radial coordinate when
\begin{equation}
z' = \frac{t_*}{2}\ .
\end{equation}
The corresponding microscopic radial scale is therefore
\begin{equation}
    z_{\rm cap} \sim \left|\frac{t_*}{2}\right| \simeq \frac{1.4}{J}\ .
\end{equation}
This scale should be interpreted as the location of the UV cap of the bilocal throat. In the regime $z,z'\gg z_{\rm cap}$, the SYK two-point function is approximately conformal and the bilocal dynamics is well described by the AdS$_2$-like throat. The conformal description should not be continued to $z,z'=0$. Instead, the throat ends at the microscopic scale $z_{\rm cap}$, where the infrared radial solution must be matched onto the short-distance behavior of the exact theory. This gives a simple estimate of the proper length of the throat. In the AdS$_2$-like region the effective metric takes the form
\begin{equation}
ds^2 \simeq \frac{-dT^2+dz^2}{z^2}\ .
\end{equation}
A fluctuation of center-of-mass frequency\footnote{Here $\Omega$ is the frequency conjugate to the average bilocal time $T=(t_1+t_2)/2$.} $\Omega$ probes radial distances of order
\begin{equation}
z_{\rm IR}\sim |\Omega|^{-1}\ .
\end{equation}
Therefore the portion of the throat accessible to this fluctuation has proper length
\begin{equation}
\ell_{\rm throat}(\Omega)\, \sim\, \int_{z_{\rm cap}}^{z_{\rm IR}} \frac{dz}{z}\, =\, \log\frac{z_{\rm IR}}{z_{\rm cap}}\, \simeq\, \log\frac{J}{1.4|\Omega|}\ .
\end{equation}

\paragraph{Relation to the black hole singularity in holography:} In the holographic context, singularities of the thermal two-point function outside the fundamental thermal strip have been argued to be the imprints of the black hole singularity. Evidence for this has come  from several fronts: thermal OPE analysis~\cite{Ceplak:2024bja,Ceplak:2025dds}, WKB asymptotics of the spectral density~\cite{Afkhami-Jeddi:2025wra,Jia:2025jbi, Giombi:2026kdz, Jia:2026ryl}, asymptotics of quasinormal modes~\cite{Festuccia:2005pi,Dodelson:2025jff}, etc. The local model arising from our analysis suggests that the spectral function of the SYK model perhaps has an asymptotic expansion (generalizing~\eqref{WKB}) 
\begin{equation}
\rho(\omega) \sim A\,\omega^\alpha\, e^{-\tau_* \omega}+ B\, \omega^\gamma\, e^{-\tau_c \,\omega} + \cdots\,,
\end{equation}
where $\tau_* \in \mathbb{R}$ and $\tau_c \in \mathbb{C}$ are the leading and subleading singularities identified in the complex time domain. We have only indicated the leading contribution to the determinant around each asymptotic piece. The Fourier transform leads to the poles in the complex time domain. This behavior is qualitatively similar to what is observed in the holographic correlators. While we have attempted to interpret the leading singularity using the kinematic space picture above, this does not as yet account for the subleading piece. It would be interesting to develop this further.

\paragraph{Applications to other models:} The techniques we developed in this paper ought to apply, in essentially unchanged form, to other large-$N$ $(0+1)$-dimensional models whose two-point functions satisfy closed-form Schwinger-Dyson equations. In particular, these include models with bosons, such as the supersymmetric cousins of the SYK model~\cite{Fu:2016vas}, and other multi-species models such as the ones analyzed in~\cite{Marcus:2018tsr,Marcus:2021ilr}. One can broadly port the technology to any melonic class of models. Of particular interest, would be the large-$N$ tensor models~\cite{Klebanov:2018fzb}, and higher-dimensional disordered models~\cite{Murugan:2017eto,Chang:2021fmd}, and models that interpolate from integrable to strong coupling limits~\cite{Peng:2018zap,Chang:2021wbx}. We hope to report on some of these in the future.

\section*{Acknowledgements}

We would like to thank I.~Araya, N.~\v{C}eplak, C.~Esper, V.~Hubeny, H.~F.~Jia, Y.~Jia, M.~Kulaxizi, H.~Liu, S.~Valach for useful discussions.
The work of I.B., I.G., E.H., A.P was supported in part by Taighde Éireann – Research Ireland under Grants SFI-22/FFP-P/11444 and 22/EPSRC/3832.
C.C.~is supported by the NSFC Grant No.~12575075. M.R.~was supported by U.S.~Department of Energy grant DE-SC0009999 and by funds from the University of California. C.C., A.P. and M.R. thank Aspen Center for Physics, which is supported by National Science Foundation grant PHY-2210452, for hospitality.

\appendix

\section{Two-point functions at finite temperature}
\label{A:Two-point functions at finite temperature}

In this appendix, we collect the definitions and properties of various finite-temperature two-point functions that appear in the main text, for the convenience of the reader. We denote Lorentzian time by $t$ and Euclidean time by $\tau = it$. The Hamiltonian of the system is denoted by $H$ and operators in the Heisenberg picture read
\begin{equation}\label{Heisenberg-evolution}
    \mathcal{O}(t) = e^{i H t}\, \mathcal{O}(0)\, e^{-i H t} = e^{i\mathcal{L}t}\mathcal{O}(0)\,,
\end{equation}
where $\mathcal{L} = \text{ad}_H$ is the ‘Liouvillian superoperator'. Throughout this appendix, we assume that $\mathcal{O}$ is a fermionic field.

\subsection{Green's functions}

Let us assume that the system resides at a finite temperature $\beta^{-1}>0$. The time-ordered Euclidean propagator is defined as
\begin{equation}\label{time-Euclidean-propagator-def}
    G(\tau) \equiv \langle\mathcal{O}(\tau)\mathcal{O}(0)\rangle_\beta\, \theta(\tau) - \langle\mathcal{O}(0)\mathcal{O}(\tau)\rangle_\beta\, \theta(-\tau)\ .
\end{equation}
where the above thermal expectation value is given by
\begin{equation}
   \langle \mathcal{O}(\tau)\mathcal{O}(0) \rangle_\beta \equiv \frac{\text{tr} \left(\mathcal{O}(\tau) \mathcal{O}(0) e^{-\beta H}\right)}{\text{tr}\left(e^{-\beta H}\right)}\ .
\end{equation}
Different analytic continuations of the Euclidean time-ordered propagator to Lorentzian time give the greater and lesser Wightman functions,
\begin{equation}\label{G>andG<}
    \begin{split}
        G^> (t) &\equiv \lim_{\varepsilon\to 0^+}G(\tau \to it +\varepsilon) =  \frac{\text{tr} \left(\mathcal{O}(t) \mathcal{O}(0) e^{-\beta H}\right)}{\text{tr}\left(e^{-\beta H}\right)}\,, \\ 
        G^< (t) &\equiv -\lim_{\varepsilon\to 0^+}G(\tau \to it -\varepsilon) = \frac{\text{tr} \left(\mathcal{O}(0) \mathcal{O}(t) e^{-\beta H}\right)}{\text{tr}\left(e^{-\beta H}\right)}\ .
    \end{split}
\end{equation}
The retarded  Green's function is then given by
\begin{equation}
    G_R(t) \equiv \theta(t) \frac{\text{tr} \left(\{\mathcal{O}(t),\mathcal{O}(0)\} e^{-\beta H}\right)}{\text{tr}\left(e^{-\beta H}\right)} = \theta(t) \left( G^>(t) + G^<(t) \right)\ .
\end{equation}
The two-sided Wightman function is defined by shifting the origin of $G^>(t)$ to the middle of the strip of analyticity
\begin{equation}\label{G12-definition}
    G_{12}(t) \equiv G^>\left(t-\frac{i\beta}{2}\right)\ .
\end{equation}
Sometimes, especially to make contact with related works \cite{viswanath_mueller_1994, Parker:2018yvk, Dodelson:2024atp}, we expand the two-point function in moments,
\begin{equation}\label{moments}
    G^>(t) \equiv \sum_{n=0}^\infty \frac{(-it)^n}{n!} \mu_n\,, \qquad \mu_n = \frac{\text{tr}\left(\mathcal{O}(0) \mathcal{L}^n \mathcal{O}(0)e^{-\beta H}\right)}{\text{tr}(e^{-\beta H})}\ .
\end{equation}
The second equation comes by Taylor-expanding the Heisenberg evolution \eqref{Heisenberg-evolution}. 

Switching to momentum space, the Fourier transform of $G_R(t)$ allows us to construct the spectral function $\rho(\omega)$, which plays an important role in our analysis
\begin{equation}\label{spectral, wightman}
    \rho(\omega) \equiv 2\text{Re} \, G_R(\omega) = \left(1+e^{-\beta\omega}\right) G^>(\omega) = G^>(\omega) + G^<(\omega)\ .
\end{equation}
Note that $G_R(\omega)$ is analytic in the upper half-plane Im$(\omega)>0$. The second and third parts of equation \eqref{spectral, wightman} are obtained using the KMS condition, as detailed below. The spectral function leads to the time-ordered Euclidean propagator
\begin{equation}\label{freq-Euclidean-propagator-def}
     G(\omega_n) = \int \frac{d\omega'}{2\pi} 
    \frac{\rho(\omega')}{\omega'-i\omega_n} 
\end{equation}
Here, $\omega_n$ are the discrete fermionic Matsubara frequencies
\begin{equation}\label{fermion-frequencies}
    \omega_n = \frac{2\pi}{\beta}\left(n+\frac12\right)\,, \qquad n\in\mathbb{Z}\,,
\end{equation}
but the form \eqref{freq-Euclidean-propagator-def} allows one to analytically continue to complex frequencies as well. Alternatively, the retarded propagator may be obtained by analytic continuation of the time-ordered Euclidean one,
\begin{equation}\label{analytic-cont-GR-G}
    G_R(\omega) = -i G(-i\omega+\varepsilon)\ .
\end{equation}

\subsection{Properties}
In equilibrium, the Wightman functions are time-translation invariant and related by
\begin{equation}\label{time-invariance-Wightman}
    G^>(t) = G^<(-t)\ .
\end{equation}
One can write the KMS condition for any of the above correlation functions. We have
\begin{align}\label{KMS-Wightman}
    & G^>(t) = G^< (t+i\beta) = G^> (-t-i\beta)\,, \qquad G^>(\omega)=e^{\beta\omega}\,G^<(\omega)\ .
\end{align}
For the Euclidean time-ordered propagator, the KMS condition is 
\begin{equation}
    G(\tau) = G(\beta-\tau)\ . \label{KMS-Euclid}
\end{equation}
Additional properties of Green's functions follow if we assume that the operator $\mathcal{O}$ is Hermitian. In this case
\begin{equation}\label{Hermiticity}
G^>(-t)=\big(G^>(t)\big)^\ast\,, \qquad G^>(t-i\beta) = G^>(t)^\ast\ .
\end{equation}

\subsection{Infinite temperature}
\label{SS:Infinite temperature}

In the infinite temperature limit, various relations between different Green's functions take a particularly simple form. In particular,
\begin{align}
    & G^>(t) = G^<(t)\,,\\
    & \rho(\omega) = 2G^>(\omega)\ . \label{rho-G>-inf-temp}
\end{align}
The moment expansion \ref{moments} also simplifies. Using Hermiticity of $H$, one can show that odd moments $\mu_{2n+1}$ vanish. Then one simply has
\begin{equation}
    G^>(t) = \sum_{n=0}^\infty \frac{(it)^{2n}}{(2n)!} \mu_{2n}\ .
\end{equation}

\section{Analytic continuation of Schwinger-Dyson equations to real time}
\label{A: Analytic continuation}

In this appendix, we explain how the SD equations \eqref{SYK SDE Euclidian} may be extended to real time and retarded and Wightman Green's functions, following  \cite{Parcollet:1999itf, Maldacena:2016hyu}. In order to do this, the crucial step is to express the Euclidean correlator in terms of the spectral density. Using \eqref{spectral, wightman}, \eqref{time-Euclidean-propagator-def} and performing analytic continuation, we can write
\begin{equation}\label{spectral representation}
    G(\tau) = \int \frac{d\omega}{2\pi} e^{-\omega\tau} \frac{\rho(\omega)}{1+e^{-\beta\omega}}\ .
\end{equation}
This representation allows to analytically continue $\Sigma(\omega_n)$ to real frequencies. We substitute \eqref{spectral representation} into the second SD equation in \eqref{SYK SDE Euclidian} in frequency space 
\begin{equation}\label{self-energy-SYK}
    \Sigma(\omega_n) = \int_0^\beta d\tau\ e^{i\omega_n\tau} G(\tau)^{q-1} = \int \left[\prod_{j=1}^{q-1} \frac{d\omega^{(j)}}{2\pi}\frac{\rho\left(\omega^{(j)}\right)}{1+e^{-\beta\omega^{(j)}}}\right] \frac{1+e^{-\beta\sum_j \omega^{(j)}}}{-i\omega_n + \sum_j\omega^{(j)}}\ .
\end{equation}
In the second step, we have also performed the integral over $\tau$. To make progress, we set $\omega_n=-i\omega + \varepsilon$ and use the Laplace identity
\begin{equation}\label{Laplace-identity}
\frac{1}{ \sum_j\omega^{(j)} - \omega-i\varepsilon} = i\int_0^\infty dt\ e^{i\omega t-\varepsilon t}e^{-it\sum_j\omega^{(j)}}\,,
\end{equation}
to rewrite \eqref{self-energy-SYK} as
{\small \begin{align}
    & \Sigma(-i\omega+\varepsilon) = \int \left[\prod_{j=1}^{q-1} \frac{d\omega^{(j)}}{2\pi}\frac{\rho\left(\omega^{(j)}\right)}{1+e^{-\beta\omega^{(j)}}}\right] \frac{1+e^{-\beta\sum_j \omega^{(j)}}}{\sum_j\omega^{(j)} - \omega - i\varepsilon}\\
    & = i\, \int_0^\infty dt e^{i\omega t - \varepsilon t} \left( \int \left[\prod_{j=1}^{q-1}\frac{d\omega^{(j)}}{2\pi} \frac{ \rho\left(\omega^{(j)}\right) e^{-it \omega^{(j)}}}{1+e^{-\beta\omega^{(j)}}}\right] + \int \left[\prod_{j=1}^{q-1}\frac{d\omega^{(j)}}{2\pi} \frac{ \rho\left(\omega^{(j)}\right) e^{(-it-\beta) \omega^{(j)}}}{1+e^{-\beta\omega^{(j)}}}\right] \right)\ . \nonumber
\end{align}}
Recognizing the Wightman correlators in the integrand, we can write
{\small \begin{align}\label{self-energy continued}
    & \Sigma(-i\omega+\varepsilon) = i\, \int_0^\infty dt e^{i\omega t - \varepsilon t} \left( \left(G^>(t)\right)^{q-1} + \left(G^>(t-i\beta)\right)^{q-1} \right) \\
    & = i\, \int_0^\infty dt e^{i\omega t - \varepsilon t} \left( \left(G^>(t)\right)^{q-1} + \left(G^>(t)^\ast\right)^{q-1} \right) = 2i\, \int_0^\infty dt e^{i\omega t - \varepsilon t} \text{Re} \left[\left(G^>(t)\right)^{q-1}\right]\ . \nonumber
\end{align}}
In order to obtain a closed system of Schwinger-Dyson equations, it remains to relate the retarded and time-ordered Euclidean correlators
\begin{equation}
    G_R^{-1}(\omega) = [-iG(-i\omega+\varepsilon)]^{-1} = -i\omega + \varepsilon -i\Sigma(-i\omega + \varepsilon)\ .
\end{equation}
Putting everything together, we get
\begin{align}\label{real-time SDE}
    &G_R^{-1}(\omega) = -i\omega + \varepsilon -i\Sigma(-i\omega + \varepsilon)\,, \nonumber\\ 
    &\Sigma(-i\omega+\varepsilon) = 2i\,\int_0^\infty dt\ e^{i\omega t-\varepsilon t}\,\text{Re}\left[\big(G^>(t)\big)^{q-1}\right]\,, \\ 
    &G^>(\omega) = \frac{2}{1+e^{-\beta\omega}}\, \text{Re}\, G_R(\omega)\ .\nonumber
\end{align}

\section{Useful formulas for computations}
\label{useful formulas}

In this appendix, we collect some formulas used in the double-expansion algorithm of Section \ref{S:Finite temperature two-point function in SYK}. Fourier and inverse Fourier transforms appearing in this algorithm are evaluated using
\begin{equation}
\mathcal{F}_\beta\{\tau^n\}(i\omega_m) \equiv \int_0^\beta d\tau\,\tau^n e^{i\omega_m\tau}
= \frac{(-1)^{n+1} n!}{(i\omega_m)^{n+1}} \left[2+\sum_{k=1}^n \frac{(-i\omega_m\beta)^k}{k!}\right]\,,
\end{equation}
\begin{equation}
\mathcal{F}_\beta^{-1}\left\{\frac{1}{(i\omega_m)^{2n+1}}\right\}(\tau)
\equiv \frac{1}{\beta}\sum_{m\in\mathbb{Z}}\frac{e^{-i\omega_m\tau}}{(i\omega_m)^{2n+1}}
= -\frac{\beta^{2n}}{2(2n)!}\, E_{2n}\left(\frac{\tau}{\beta}\right)\ .
\end{equation}
In the second line, it is assumed that $0<\tau<\beta$ and the Euler polynomials $E_n(x)$ are defined through the generating function
\begin{equation}
\frac{2e^{xz}}{e^z+1} \equiv \sum_{n=0}^\infty E_n(x)\,\frac{z^n}{n!}\ .
\end{equation}
When doing computations in real time, one makes use of the Fourier transforms
\begin{equation}
F^{-1}\{\delta^{(n)}(\omega)\}(t)=\int_{-\infty}^\infty \frac{d\omega}{2\pi}\,\delta^{(n)}(\omega)\,e^{-i\omega t} = \frac{(it)^n}{2\pi}\,,
\end{equation}
\begin{equation}
F\{t^n\}(\omega) = \int_0^\infty dt\,t^n\,e^{i\omega t-\varepsilon t}
= \frac{n!}{(-i\omega+\varepsilon)^{n+1}}\ .
\end{equation}
Other useful formulas that are needed for real time expansion and double expansion algorithms read
\begin{equation}
    \text{Re} \frac{1}{(-i\omega+\varepsilon)^{2n+1}} = \frac{(-1)^{n}}{(2n)!}\,\pi\delta^{(2n)}(\omega)\,,
\end{equation}
\begin{equation}
    \frac{\delta^{(n)}(\omega)}{1+e^{-\beta\omega}}
=\frac12\,\delta^{(n)}(\omega)
-\sum_{m=1}^{\lceil n/2\rceil}\beta^{2m-1}\binom{n}{2m-1}
\frac{(2^{2m}-1)B_{2m}}{2m}\,
\delta^{(n-2m+1)}(\omega)\,,
\end{equation}
where $B_{2m}$ are Bernoulli numbers.


\bibliographystyle{JHEP}
\bibliography{bibliography} 

\end{document}